\documentclass[11pt]{article}
\usepackage{amsthm}
\usepackage{amsmath}
\usepackage{natbib}
\usepackage[colorlinks,citecolor=blue,urlcolor=blue,filecolor=blue,backref=page]{hyperref}
\usepackage{graphicx}

\usepackage{authblk}
\usepackage{amssymb}
\usepackage{psfrag,epsf}
\usepackage{enumerate}

\usepackage{url}
\usepackage{bm, bbm}
\usepackage{color}
\usepackage{subfig}
\usepackage{morefloats}
\usepackage{adjustbox}
\usepackage{makecell}
\usepackage{multirow}
\usepackage{tikz}
\usetikzlibrary{calc}

\newcommand{\iid}{\stackrel{\rm iid}{\sim}}
\newcommand{\ind}{\stackrel{\rm ind}{\sim}}

\begin{document}

\title{Seemingly Unrelated Multi-State processes: a Bayesian semiparametric approach}

\author[1]{Andrea Cremaschi}
\author[2,1,3,4]{Raffele Argiento}
\author[2,1,3,4]{Maria De Iorio}
\author[1]{Cai Shirong}
\author[2,1]{Yap Seng Chong}
\author[1,2,5]{Michael J. Meaney}
\author[1]{Michelle Z. L. Kee}

\affil[1]{Singapore Institute for Clinical Sciences, A*STAR, Singapore}
\affil[2]{Yong Loo Lin School of Medicine, National University of Singapore, Singapore}
\affil[3]{Division of Science, Yale-NUS College, Singapore}
\affil[4]{Department of Statistical Science, University College London, UK}
\affil[5]{Department of Psychiatry, McGill University, Montreal, Canada}

\date{}

\maketitle

\textbf{Abstract}: Many applications in medical statistics as well as in other fields can be described by transitions between multiple states (e.g. from health to disease) experienced by individuals over time. In this context, multi-state models are a popular statistical technique, in particular when the exact transition times are not observed. The key quantities of interest are the transition rates, capturing the instantaneous risk of moving from one state to another. The main contribution of this work is to propose a joint semiparametric model for several possibly related multi-state processes (Seemingly Unrelated Multi-State, SUMS, processes), assuming a Markov structure for the transitions over time. The dependence between different processes is captured by specifying a joint random effect distribution on the transition rates of each process. We assume a flexible random effect distribution, which allows for clustering of the individuals, overdispersion and outliers. Moreover, we employ a graph structure to describe the dependence among processes, exploiting tools from the Gaussian Graphical model literature. It is also possible to include covariate effects. We use our approach to model disease progression in mental health. Posterior inference is performed through a specially devised MCMC algorithm.


\section{Introduction}\label{sec:Intro}

Biomedical data are characterized by a high number of different variables, in many cases mostly categorical and recorded on a (nowadays often large) set of subjects. This is mainly due to the practice in clinical settings to record the absence/presence of symptoms and/or to use ordinal scales to represent disease markers. Typically, we only observe clinical variables at fixed time points (usually corresponding to follow up or hospital visits), and as such these variables are interval-censored (i.e., panel data). 
The objective of clinical studies is often to model a patient's \textit{disease progression}, as captured by the evolution over time of one or more responses of interest, e.g. representing the disease status, and associated clinical markers. A popular approach to  disease progression is to use multi-state models describing the transition of individuals among multiple states in continuous time \citep[see, for instance,][]{Cook_1999,  jackson_2011, van_2015, DeIorio_etal_2018}. In this framework, it is straightforward to include time-homogeneous covariates and time varying continuous covariates (leading to a Markov regression model).

In this work, we propose a joint modelling approach for several categorical variables evolving simultaneously through time. More in details, our approach is based on a combination of ideas from multi-state models, seemingly unrelated regression \citep{zellner1963estimators, wang2010sparse}, Gaussian Graphical models and  Product Partition Models with Covariates (PPMx) \citep{muller2011product}. In a Bayesian framework, we define a joint model for several multi-state processes, which represent the evolution of, for instance, clinical markers of interest  as in the disease progression application of Section \ref{sec:GUSTO_application}. The main idea is that the different markers provide complementary information on the underlying health status and, as such, they are regarded as stochastic processes defined on a finite state-space, evolving in continuous time according to \textit{dependent} Markov processes. We link the different Markov processes through the specification of a flexible prior distribution on the instantaneous transition rates, specifically a mixture distribution with random number of components \citep{argiento2019infinity}. In this way, we build a robust modelling strategy, which leads to covariate-driven clustering of the subjects and enables the inclusion of different types of covariates/responses in a natural and efficient way \citep{barcella2017comparative}. Each multi-state process is then, conditionally on the vector of instantaneous transition rates, independent from the other processes, resembling the seemingly unrelated regressions (SUR) setting of \cite{zellner1963estimators}. Furthermore, we allow the dependence structure between the transition rates to be encoded into a random graph, which is also object of posterior inference, as it is done in sparse SUR models \citep[SSUR,][]{wang2010sparse}. Thus, the nature of the dependence is learnt from the data, spanning from independence to full inter-dependence. As such, we refer to our model as \textit{Seemingly Unrelated Multi-State} (SUMS) processes. Briefly, the proposed model allows for: (i) multiple responses; (ii) processes with more than two states; (iii) patient- and process-specific times of observation; (iv) inclusion of mixed-type covariates; (v) covariate-driven clustering of the subjects; (vi) missing initial state information. 

One of the main advantages of our modelling strategy is that the relationship between different multi-state processes is encoded into a
graph structure. Indeed, if there is an edge linking two processes, it means that they are conditionally dependent, while the lack of an edge implies conditional independence. This gives insight into the co-regulatory mechanisms of the different processes. This is relevant in many application as often it is of interest also to identify important factors affecting disease progression, for better prognosis and therapeutic choices. Moreover, the model allows for the inclusion of time-homogeneous covariates (of any type) and time-varying continuous covariates in a regression component, for which standard variable selection techniques (e.g. shrinkage, spike and slab priors) can be employed.

The manuscript is organised as follows: Section \ref{sec:SUMS} introduces the SUMS model, by presenting how its key components - the joint multi-state model, the mixture prior with unknown number of components and the graphical structure describing the dependence among processes - interrelate, as well as the  specifically designed MCMC algorithm. Section \ref{sec:GUSTO_application} presents an application of the model to the analysis of mental health indicators obtained from the GUSTO cohort study. Section \ref{sec:Conclusions} concludes the work. In Supplemetary Material we include a detailed description of the algorithm and of the GUSTO dataset, a sensitivity analysis and a simulation study, as well as further results from the analysis of the GUSTO data.

\section{SUMS: Seemingly Unrelated Multi-State Processes}\label{sec:SUMS}

\subsection{Modelling of multi-state processes}\label{sec:modelling-of-multi-state-processes}
Multi-state models can be used to describe how an individual moves between a set of states in continuous time. In this work, we focus on multi-state processes for panel data, where the states of several processes are observed only at certain time points, and their exact transition times are not known. For each $h = 1, \dots, p$, let $\{ Y^{(h)}(t), t \in \mathbb{R}^+ \}$ be  a continuous time Markov process,  where $Y^{(h)}(t)$ represents the state of the $h-$th process over time, with state-space $\mathcal{S}^{(h)} = \{1, \dots, d^{(h)}\}$ of dimension $d^{(h)}$, i.e. $Y^{(h)}(t) \in \mathcal{S}^{(h)}$; the elements of $\mathcal{S}^{(h)}$ represent the states that the $h$-th process can visit between transitions. The exact times of transition of the processes $Y^{(h)}(t)$ are not known, but in applications the processes are observed on a discrete set of time points, $\bm t_i^{(h)} = \left(t_{i1}^{(h)},\ldots, t_{in^{(h)}_i}^{(h)}\right)$, where $n^{(h)}_i$ denotes the number of observed time points for the $i$-th individual and $h$-th process. Notice that the times of observation and their number are both process- and subject-specific. We indicate with $Y^{(h)}_{ij} = Y^{(h)}(t_{ij}^{(h)})$ the value of the $h$-th process $Y^{(h)}(t)$ at the $j$-th observed time $t_{ij}^{(h)}$ for the $i$-th subject. Hence, for each subject $i = 1, \dots, N$, we observe the random vector $\bm Y^{(h)}_i = \left(Y^{(h)}_{i1}, \dots, Y^{(h)}_{in^{(h)}_i}\right)$, whose joint distribution is the finite-dimensional law of the process $Y^{(h)}(t)$ at the times of observation, for $h = 1, \dots, p$. The aim of this work is to jointly model the processes $Y^{(h)}(t)$, capturing their time evolution and possible dependencies. For each process, we assume that the Markov property holds, i.e. conditionally on current and past events, future transitions only depend on the current state. The probability law of the $h$-th process $Y^{(h)}(t)$ is assigned via the matrix of instantaneous transition rates $\bm Q^{(h)}(t) =[\lambda^{(h)}(r,s;t)]_{r,s} $, which is also time-dependent, and whose elements are the instantaneous transition rates $\lambda^{(h)}(r,s;t)$ with $r, s \in \mathcal{S}^{(h)}$. In what follows, the vector $\bm \lambda^{(h)}(t) = \{\lambda^{(h)}(r,s;t): r, s \in \mathcal{S}^{(h)}, r \neq s\}$, of dimension $D_p = \sum_{h=1}^{p}d^{(h)}(d^{(h)}-1)$, indicates the off-diagonal transition rates of the matrix $\bm Q^{(h)}(t)$ at time $t > 0$, concatenated by row from top to bottom. For simplicity, we indicate a transition of the $h$-th process between different states of $\mathcal{S}^{(h)}$ with the notation $r \rightarrow s$. The instantaneous transition rates can be made covariate-dependent by specifying a Cox proportional hazard model. This allows the inclusion of both time-homogeneous covariates as well as time-varying continuous covariates. Alternatively, a  semi-proportional intensity model can be easily specified for the covariates as in \cite{kim2012bayesian}.  Note that the decision of including either type of covariates is process-specific. The time-homogeneous covariates are straightforwardly incorporated in the model, and we denote them here by $\bm X^{(h)}_i = \left(X^{(h)}_{i1}, \dots, X^{(h)}_{ig^{(h)}}\right)$, for the $i$-th individual and $h$-th process. On the other hand, the time-varying continuous ones, denoted by $\bm Z^{(h)}_i(t) = \left(Z^{(h)}_{i1}(t), \dots, Z^{(h)}_{iq^{(h)}}(t)\right)$, require additional assumptions. They are usually included by assuming a piece-wise constant effect over each interval of observations \citep{andersen2012statistical}, or by modelling them as longitudinal processes, linking their distribution to the ones of the multi-state processes via the inclusion of suitable random effects \citep{ferrer2016joint}. The first option has a clear computational advantage, while the latter has the potential to yield better inference on the overall disease progression. In the application presented in Section \ref{sec:GUSTO_application} we are not provided with any time-varying continuous covariates. However, the code provided with this manuscript allows for the implementation of the first method. This assumption leads to a piecewise constant model for the instantaneous transition rates $\lambda^{(h)}_{ij}(r,s)$, with $r, s \in \mathcal{S}^{(h)}$, and for the matrix $\bm Q^{(h)}_{ij} := \bm Q^{(h)}_i(t^{(h)}_{ij})$, for $j = 1, \dots, n^{(h)}_i$, $i = 1, \dots, N$ and $h = 1, \dots, p$. The model for the instantaneous $\log$-transition rates for $i = 1, \dots, N$ is then:
\begin{equation}\label{eq:Lambdas_Y}
\log\left(\lambda^{(h)}_{ij}(r,s)\right) = \phi^{(h)}_i(r,s) + \bm X^{(h)}_i \bm \beta^{(h)}_{rs} + \bm Z^{(h)}_{ij} \bm \gamma^{(h)}_{rs}, \quad j = 1, \dots, n^{(h)}_i, h = 1, \dots, p
\end{equation}
where $\phi^{(h)}_i(r,s)$ represents the baseline transition rate (on a log scale) of a transition $r \rightarrow s$. The parameters $\bm \beta^{(h)}_{rs} \in \mathbb{R}^{g^{(h)}}$ and $\bm \gamma^{(h)}_{rs} \in \mathbb{R}^{q^{(h)}}$ are the vectors of regression coefficients for the $h$-th process and the $r\rightarrow s$ transition.

Let $\epsilon^{(h)}_{ij} = t^{(h)}_{ij} - t^{(h)}_{ij-1}$ indicate the length of the $j$-th time interval, for $j = 2, \dots, n^{(h)}_i$, $i = 1, \dots, N$ and $h = 1, \dots, p$. Thanks to the piecewise constant assumption, the Chapman-Kolmogorov equations can be solved to obtain the interval-specific transition probabilities, $\bm p_{ij}^{(h)} (\bm \lambda^{(h)}_{ij} ,\epsilon^{(h)}_{ij}) = \left\{p_{ij}^{(h)} (r, s; \bm \lambda^{(h)}_{ij} ,\epsilon^{(h)}_{ij}): r, s \in \mathcal{S}^{(h)}\right\} $, for the vector of random variables $\bm Y^{(h)}_i$ \citep[see][]{Ross_1996}.
When $d^{(h)} = 2$, closed-form solutions are readily available \citep{cox1977theory}, while problems involving more than three states are usually tackled numerically \citep{moler2003nineteen}. 
It can be shown that for each process $h$ a unique stationary distribution exists \citep{grimmet1992probability}, and we denote it by $\bm \pi^{(h)}_{ij}\left(k; \bm \lambda^{(h)}_{ij}\right)$, with $k \in \mathcal{S}^{(h)}$, highlighting the fact that these are functions of the subject-specific instantaneous transition rates \citep[see][for details]{Ross_1996}. 
The stationary distribution can be used as marginal distribution for modelling the state of the processes at time $j = 1$, considering the vectors of instantaneous transition rates $\bm \lambda^{(h)}_{i1}$, in contrast to the general practice in multi-state modelling of specifying the model conditionally on the state at the first time of observation. This is important, as it allows Bayesian imputation of missing observations at time one, since they are treated as unknown parameters in the model. This aspect is particularly useful in our application, where the initial time presents a non-negligible missing rate. 

The specification of the process- and subject-specific transition probabilities, together with the existence of the stationary distribution, leads to the joint likelihood for the vector of observed states $\bm Y^{(h)}_i$ for $i = 1, \dots, N$ and $h = 1, \dots, p$, as follows:
\begin{equation}\label{eq:likelihood_Y}
p\left(\bm Y \mid \bm \lambda^Y \right) = \prod_{i = 1}^N\prod_{h = 1}^{p}\prod_{j = 2}^{n^{(h)}_i} \left( p^{(h)}_{ij} \left( Y^{(h)}_{ij-1}, Y^{(h)}_{ij}; \bm \lambda^{(h)}_{ij},\epsilon^{(h)}_{ij}\right) \right) \pi^{(h)}_{i1}\left(Y^{(h)}_{i1}; \bm \lambda^{(h)}_{i1}\right)
\end{equation}	
where $\bm Y$ and $\bm \lambda^Y$ indicate the multi-dimensional arrays containing the observation vectors $\bm Y^{(h)}_i$ and the instantaneous transition rate vectors $\bm \lambda^{(h)}_i$, while  $p_{ij}^{(h)}$ denotes the transition probabilities and $\pi^{(h)}_{i1}$ is the (stationary) distribution at time one. 

\subsection{Random effect distribution and relationship with SUR}\label{sec:BNP_Model}

Consider the vector of $\log$-baseline transition rates  for the $h$-th process $\bm \phi^{(h)}_i = \{\phi^{(h)}_i(r,s): r \rightarrow s\}$, for $h = 1, \dots, p$ and for each subject $i$, and let $\bm \phi_i = (\bm \phi^{(1)}_i, \dots, \bm \phi^{(p)}_i)$ be the vector containing the $\log$-baseline transition rates of all the $p$ processes. To capture the inter-individual heterogeneity and allow for clustering of the subjects, we choose as random effect distribution for $\bm \phi_1, \dots, \bm \phi_N$ a mixture prior with random number of components, where the distribution of the weights is given by the normalization of a finite point process, as proposed by \cite{argiento2019infinity}. This approach has several advantages, allowing for flexible modelling of the weights in the mixture as well as efficient posterior computations (e.g., as compared to traditional reversible jump algorithms for mixture models). In more details, we assume the following mixture prior:
\begin{align}\label{eq:NIFPP_model}
	\bm \phi_i &= \bm \phi^\star_{c_i}, \quad i = 1, \dots, N \nonumber \\
	\bm \phi^\star_1, \dots, \bm \phi^\star_M \mid M &\iid P_0(\bm \phi^\star \mid \bm \theta) \nonumber \\
	\mathbb{P}(c_i = m)& \propto S_m, \quad i = 1, \dots, N \\
	S_1, \dots, S_M & \iid \text{Gamma}(\gamma_S, 1) \nonumber \\
	M - 1 & \sim \text{Poi}(\Lambda) \nonumber 
\end{align}
where we denote by $\text{Gamma}(a, b)$ the Gamma distribution with mean $a/b$, and by $\text{Poi}(\Lambda)$ the Poisson distribution with mean $\Lambda$. The variables $\bm c = (c_1, \dots, c_N)$ indicate the component allocations of the subjects and their corresponding prior probabilities are proportional to the unnormalized weights $\bm S = \left(S_1, \dots, S_M\right)$. As shown by \cite{argiento2019infinity}, posterior computation is greatly simplified via the introduction of a latent variable, conditionally on which the unnormalized weights of the mixture become independent. This computational trick is borrowed from the Bayesian nonparametric literature \citep{jamesetal09}. 
Finally, the vectors $\bm \phi^\star_1, \dots, \bm \phi^\star_M$ are a finite sequence of locations for the mixture distribution and are, conditionally on the number of components $M$, i.i.d. from the base measure $P_0$. The specification of a joint random effect distribution for $\bm \phi_1, \dots, \bm \phi_N$ in model \eqref{eq:NIFPP_model} and the choice of $P_0$ are crucial in our modelling strategy, as it will be shown in Section \ref{sec:GGM}, since this allows inference on the shared dependence structure among the components of the vectors $\bm \phi^\star_m$, for $m = 1, \dots, M$ and, consequently, on 
 the dependence structure
 among the $p$ different processes. As an alternative, a Bayesian nonparametric prior could have been specified as random effect distribution such as the Dirichlet process \citep{DeIorio_etal_2018} and the beta-Dirichlet process prior  \citep{kim2012bayesian}, or, taking a complete different approach, flexible modelling of the baseline transition intensities   can be achieved using penalised splines \citep{kneib2008bayesian}. 
 
 Our modelling approach resembles the one underlying the SUR framework of \cite{zellner1963estimators}, where $p$ different regression models are linked by specifying a joint error distribution, usually multivariate normal. The SUR methodology is one of the main techniques for handling multiple responses and offers a way to share information between models which are \textit{seemingly} unrelated, since they describe different data-generating processes. However, since these are observed for the same set of subjects and measurements are taken on often related processes, the study of their interdependency is of great interest in most applications. For this reason SUR-type models have gained vast popularity in different fields, such as Phenomics \citep{houle2010phenomics, banterle2018sparse}. In our application, for instance, where we deal with several processes associated to different aspects of maternal mental health (e.g., depression, anxiety, sleep quality), it is important to understand the relationships between such processes in order to have a broader view of the phenomenon under study. As in the SUR framework, in our context each process is modelled by its own \textit{seemingly unrelated} multi-state Markov process, but then they are \textit{related} through the joint random effect distribution on $\bm \phi = \left(\bm \phi_1, \dots, \bm \phi_N\right)$.  Motivated by this parallelism, we name the proposed model as Seemingly Unrelated Multi-State (SUMS) processes.

\subsection{Gaussian Graphical model}\label{sec:GGM}
We use tools from the Gaussian Graphical models literature to describe  the dependence 
among the $p$ processes. Referring to model \eqref{eq:NIFPP_model}, we assume that $\bm \phi^\star_1, \dots, \bm \phi^\star_M \mid M \iid P_0 = \text{N}\left(\bm \mu , \bm \Omega_G\right)$, where the key modelling feature is the specification of the prior on the precision matrix $\bm \Omega_G$ conditional on a graph $G$, which captures the conditional dependence structure among the $\log$-baseline transition rates. The novelty of our modelling strategy is that $G$ is modelled conditionally on another random graph $G_0$, which characterizes the dependence structure among processes and is one of the main aspects of our inference. 

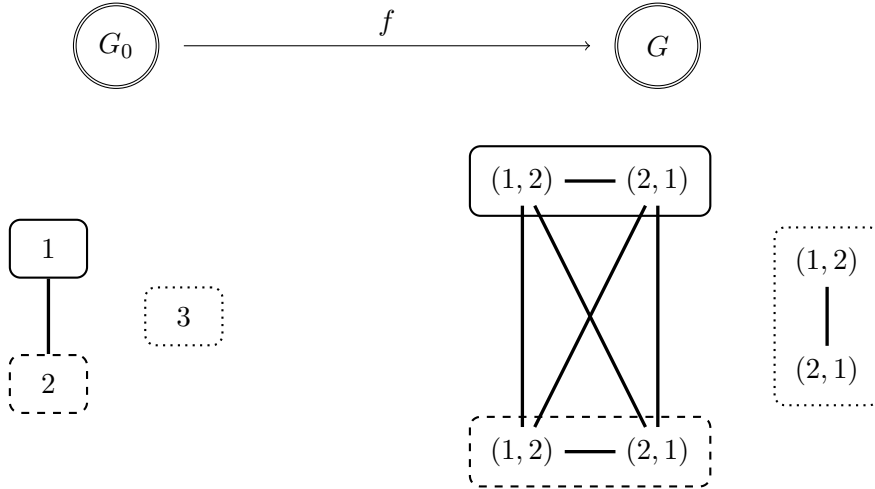
\begin{figure}
	\centering
	\begin{tikzpicture}[scale=1.8, auto, swap,
		block1/.style={
		rectangle,
		draw=black,
		thick,
		fill=white,
		text width=2em,
		align=center,
		rounded corners,
		minimum height=2em
		},
		block2/.style={
		rectangle,
		draw=black,
		thick, dashed,
		fill=white,
		text width=2em,
		align=center,
		rounded corners,
		minimum height=2em
		},
		block3/.style={
		rectangle,
		draw=black,
		thick, dotted,
		fill=white,
		text width=2em,
		align=center,
		rounded corners,
		minimum height=2em
		}]
	
		\node[draw=black,double,inner sep=1ex,circle,minimum size=1.1cm] (G0) at (0,2.5) {$G_0$};
		\node[block1] (Y1) at (-0.5,1) {$1$};
		\node[block2] (Y2) at (-0.5,0) {$2$};
		\node[block3] (Y3) at (0.5,0.5) {$3$};
		
		\node[draw=black,double,inner sep=1ex,circle,minimum size=1.1cm] (G) at (4,2.5) {$G$};
		\node (l112) at (3,1.25+0.25) {$(1,2)$};
		\node (l121) at (4,1.25+0.25) {$(2,1)$};
		\node (l212) at (3,-0.25-0.25) {$(1,2)$};
		\node (l221) at (4,-0.25-0.25) {$(2,1)$};
		\node (l312) at (5.25,0.5+0.4) {$(1,2)$};
		\node (l321) at (5.25,0.5-0.4) {$(2,1)$};
		
		\path[line width = 1.25pt] (Y1) edge node {} (Y2);
		\path[line width = 1.25pt] (l112) edge node {} (l121);
		\path[line width = 1.25pt] (l212) edge node {} (l221);
		\path[line width = 1.25pt] (l312) edge node {} (l321);
		\path[line width = 1.25pt] (l112) edge node {} (l212);
		\path[line width = 1.25pt] (l112) edge node {} (l221);
		\path[line width = 1.25pt] (l121) edge node {} (l212);
		\path[line width = 1.25pt] (l121) edge node {} (l221);
		
		\draw[->] (0.5,2.5) -- node[above]{$f$} (3.5,2.5);
		
		\draw[thick, rounded corners] ($(l112.north west)+(-0.075,0.075)$) rectangle ($(l121.south east)+(0.075,-0.075)$);
		\draw[thick, rounded corners, dashed] ($(l212.north west)+(-0.075,0.075)$) rectangle ($(l221.south east)+(0.075,-0.075)$);		
		\draw[thick, rounded corners, dotted] ($(l312.north west)+(-0.075,0.075)$) rectangle ($(l321.south east)+(0.075,-0.075)$);
				
	\end{tikzpicture}
	\caption{Example of graphical structures underlying the SUMS processes: the graph $G_0$ describes the conditional dependence between the processes in $V_0 = \{1, 2, 3\}$, while the graph $G$ is obtained as a deterministic function from $G_0$ by connecting the edges of the corresponding transitions between states 1 and 2 of each process.} \label{fig:tikz_G_G0}
\end{figure}

In details, consider the graph $G_0 = (V_0, E_0)$, defined over the set of nodes $V_0 = \{1, \dots, p\}$, i.e. each node in the graph corresponds to a multi-state process
$Y^{(h)}(t)$. The edge set $E_0$ is formed of the pairs $E_0 \subseteq  \{(h,k)\in V_0\times V_0: h <k  \}$ such that an edge exists between nodes $h, k \in V_0$. We consider only simple graphs, i.e. undirected graphs, without self-loops nor multiple edges. As mentioned earlier, to introduce dependence among the elements of the vector $\bm \phi^\star_m = (\bm \phi^{\star, (1)}_m, \dots, \bm \phi^{\star, (p)}_m)$, with $\bm \phi^{\star, (h)}_m = \{\phi^{\star, (h)}_m(r,s): r \rightarrow s\}$ for $h = 1, \dots, p$, we define a second graph $G$ whose structure is determined by $G_0$. In particular, we let $G = (V, E)$ be the graph whose nodes are the indices of the vector $\bm \phi^\star_m$, i.e. $V = \{1, \dots, D_p\}$, with $D_p = \sum_{h \in V_0} d^{(h)}(d^{(h)} - 1)$. $G$ is a deterministic function of $G_0$ specified as follows. First, whatever the form of $G_0$, there exists an edge in $G$ between transition rates of the same process. Therefore, an empty graph $G_0$ corresponds to a graph $G$ with $p$ cliques, one for each process. Second, if there is an edge between nodes $h$ and $k$ in $G_0$ (i.e., $(h,k) \in E_0$), then there is and edge between all the possible pairs of elements of $\bm \phi^{\star, (h)}_m$ with those of $\bm \phi^{\star, (k)}_m$. An illustration for the case of three binary processes is given in Figure \ref{fig:tikz_G_G0}. We write $G = f(G_0)$, $f$ being the transformation  described above. Note that $f$ is bijective and, as such, the specification of a prior on $G_0$ implies  a prior on $G$. This construction is advantageous in terms of dimension reduction, as the dimension of the graph space where $G_0$ is defined can be significantly smaller than the one of $G$, leading to more efficient exploration of the posterior space.

Following the literature on GGMs, the conditional independence structure of the multivariate Gaussian vectors $\bm \phi^\star_m \sim \text{N}(\bm \mu, \bm \Omega_G)$, for $m = 1, \dots, M$, is described by constraining the elements of the precision matrix $\bm \Omega_G$ \citep{Dempster}. Namely, two elements of the vector $\bm \phi^\star_m$ are, conditionally on the others, independent if and only if there is a zero in the corresponding entry of the precision matrix $\bm \Omega_G$. Since $G$ is a deterministic function of $G_0$, it is the latter that encodes the conditional independence structure of the vectors $\bm \phi^\star_m$, for $m = 1, \dots, M$ (see Figure \ref{fig:tikz_G_G0}). The standard conjugate prior for the precision matrix $\bm \Omega_G$ is the G-Wishart distribution, specified conditionally on the graph structure $G$ \citep{Roverato2002}. The last component needed to fully specify this part of the model, is the prior distribution for the graph $G$. We do not assign this prior directly, but rather it is inherited by the prior we choose for the graph $G_0$:
$$
\pi(G_0 \mid \eta) \propto \eta^{|E_0|} (1 - \eta)^{\binom{p}{2} - |E_0|}, \quad \eta \in (0,1)
$$
where $|E_0|$ is the number of edges in graph $G_0$ (i.e., the size of $E_0$), while $\binom{p}{2}$ is the number of possible graphs with nodes $V_0 = \{1, \ldots, p\}$. This prior is equivalent to assuming a Bernoulli prior with probability of success (here inclusion) $\eta$ on each edge of the graph $G_0$, independently across edges. Small values of $\eta$ favour sparser graphs \citep{armstrong2009}. Finally, we point out that, while the prior for the graph $G_0$ is defined over all possible graphs, including the non-decomposable ones, the resulting prior distribution on $G$ is defined on a restricted space due to the clique constraints imposed on the  transitions of the same process which need to  be  fully connected. 

\subsection{Relationship with PPMx models}\label{sec:relationship-with-ppmx-models}

The SUMS model has wide applicability in biomedical research as some processes can be regarded as responses and some as covariates. Indeed, time-varying categorical covariates are very common in the field of medical research, for instance in association with the monitoring of a patient's disease status over time. 
In the application to disease progression in Section \ref{sec:GUSTO_application}, some processes represent mental health outcomes of interest, while others correspond to categorical clinical markers, and the goal of the analysis is to model the joint evolution of outcomes and clinical factors to determine how the symptom variables influence the disease course. 
Handling of time-varying multivariate categorical information can be problematic in several applications and the SUMS approach provide a natural framework to deal with this problem. In a Bayesian framework, \cite{DeIorio_etal_2018} discuss possible solutions and propose an approach based on a latent health function borrowing ideas from Item Response Theory \citep{thissen2009item}. The latter approach, although computationally efficient, does not allow for a direct quantification of the covariate effect on the clinical response of interest and it may lead to identifiability problems. A simpler and more common approach to deal with time-varying categorical covariates is to introduce appropriate dummy variables, considerably  increasing  the number of parameters to be estimated, resulting in slower computations and lower effectiveness in high dimensional problems.
Another computational effective solution is to summarize the covariates into an  often arbitrary  time-varying  score, but at the cost of losing information and interpretability. 

When some of the multi-state processes can be seen as covariates, then the SUMS model has interesting connections with Product Partition Models with Covariates (PPMx), very popular in the Bayesian nonparametric literature \cite{muller1996bayesian, muller2011product}. Indeed, our approach provides a flexible and robust modelling strategy, which leads to covariate-driven clustering of the subjects and enables the inclusion of different types of covariates/responses in a natural and efficient way \citep{barcella2017comparative}. We now clarify the relationship between SUMS and PPMx. The main modelling idea behind the PPMx is to include covariate information into the partition model (e.g., into a Dirichlet Process Mixture model framework) by treating each covariate as a random variable \citep{muller1996bayesian, muller2011product, barcella2017comparative}. In the SUMS approach we can consider a set of processes $(Y^{(1)}(t), \dots, Y^{(p_Y)}(t))$ as responses and another set, denoted by $(H^{(1)}(t), \dots, H^{(p_H)}(t))$, as explanatory factors. Then Eq.\eqref{eq:likelihood_Y} specifies a suitable probability model for the joint vector of processes $\{(Y^{(h)}(t), H^{(l)}(t)); h=1, \ldots, p_Y; l=1,\ldots,p_H; t \in \mathbb{R}^+ \}$, with $p_Y+p_H=p$. Let $\bm c$ be the set of allocation variables introduced in \eqref{eq:NIFPP_model}, and let $\rho_N$ be the partition of the indices $\{1, \dots, N\}$ induced by $\bm c$. We indicate by $C_j$ the set of indices belonging to the $j$-th cluster, i.e. $C_j = \{i \in \{1, \dots, N\} \mid c_i = j\}$, and thus a partition with $K_N$ clusters corresponds to $\rho_N = \{C_1, \dots, C_{K_N}\}$. In the PPMx framework, the SUMS model induces a prior on the partition $\rho_N$ which depends on the covariates $\bm H$:
\begin{equation}\label{eq:rho_H_1}
	p\left(\rho_N \mid \bm H \right) = V(N, K_N) \prod_{j = 1}^{K_N} \mathcal{C}(C_j) \mathcal{G}(\bm H^\star_j)
\end{equation}
where $\mathcal{C}(C_j)$ is the \textit{cohesion}, i.e. a function of the $j$-th cluster $C_j$, $\mathcal{G}(\bm H^\star_j)$ is the \textit{similarity}, i.e. a function of the array of covariates corresponding to the subjects in cluster $j$, denoted as $\bm H^\star_j := \{H_i : i \in C_j\}$, for $j = 1, \dots, K_N$, and $V(N, K_N)$ is a constant depending only on the sample size $N$ and the number of clusters $K_N$. The cohesion function $\mathcal{C}$ expresses prior information about the partition, such as the average size of a cluster, while the similarity function $\mathcal{G}$ captures the contribution of the covariates to the clustering structure. The presence of $\mathcal{G}$ in \eqref{eq:rho_H_1} allows subjects with similar covariates to be more likely assigned to the same cluster. Under our modelling assumptions it can be shown that \eqref{eq:rho_H_1} is given by: 
\begin{gather}
	p\left(\rho_N \mid \bm H \right) \propto p\left( \rho_N \right) p\left(\bm H \mid \rho_N \right) = \nonumber \\
	V(N, K_N) \prod_{j = 1}^{K_N} \frac{\Gamma(\gamma_S + n_j)}{\Gamma(\gamma_S)} \int \left( \prod_{i \in C_j}  p\left(\bm H_i \mid \bm \lambda^H_i \right) \right) P_0(d\bm \phi^{\star, (p_Y + 1)}_j, \dots, d\bm \phi^{\star, (p)}_j) \label{eq:rho_H_2}
\end{gather}
See Supplementary Material Section 1 for a proof. The proposed model induces the similarity function $\mathcal{G}(\bm H^\star_j) = \int \prod_{i \in C_j}  p\left(\bm H_i \mid \bm \lambda^H_i \right) P_0(d\bm \phi^{\star, (p_Y + 1)}_j, \dots, d\bm \phi^{\star, (p)}_j)$, for  $j = 1, \dots, K_N$. The similarity function $\mathcal{G}$ is not known in closed form, differently from the common  PPMx specification, where the similarity function is usually obtained from a conjugate model for the covariates vector via marginalization to simplify computations. In the proposed approach, the evaluation of \eqref{eq:rho_H_2} would require an expensive numerical approximation. For this reason, we resort to a conditional MCMC algorithm analogous to the one proposed by \cite{argiento2019infinity}, not requiring the evaluation of the integral in \eqref{eq:rho_H_2}.

\subsection{MCMC Algorithm}\label{sec:mcmc-algorithm}
Posterior inference is performed through a MCMC algorithm, described in details in Supplementary Material Section 2. The numerous non-conjugate updates required by the proposed model are tackled using adaptive Metropolis-Hastings sampling schemes \citep{haario2001adaptive, atchade2005adaptive}, which need an additional short burn-in period. Additionally, inference under the proposed model is challenging given the presence of the graphs $G$ and $G_0$. We adopt the birth-and-death approach of \cite{mohammadi2015bayesian}, and extend their algorithm to accommodate for MCMC moves on cliques instead of single edges, recalling that each edge in $G_0$ corresponds to a clique in $G$ through the map $f$. Indeed, the original algorithm of \cite{mohammadi2015bayesian} is based on theoretical results from the GGM literature \citep[see][]{wang2012efficient}, which can be extended to our modelling setting. In Supplementary Material Section 4, we also compare the performance of our model with the approach of \cite{DeIorio_etal_2018} and with two alternative versions of the proposed model (i.e., DP and parametric versions). The results of the comparison show that the proposed model outperforms the parametric approach, as well as the nonparametric competitors in terms of clustering, leading to comparable results with respect to the estimation of regression coefficients.

\section{Application to the GUSTO study}\label{sec:GUSTO_application}
The GUSTO study \citep[Growing Up in Singapore Towards healthy Outcomes, ][]{soh2014cohort} is a longitudinal birth cohort study started in 2009 and involving Singaporean mothers and their children. The study is one of the most carefully phenotyped parent-offspring cohorts, focusing on the roles of foetal, developmental and epigenetic factors involved in early body composition as well as neuro-development. In this work we consider data on $N = 301$  mothers, followed during pre- and post-natal periods, starting from three months before childbirth. The main focus of the analysis is understanding the relationship among five psychometric indicators obtained from specific questionnaires: the Beck's Depression Inventory II \citep[BDI II,][]{Beck_etal_1961}; the Edinburgh Postnatal Depression Scale \citep[EPDS,][]{matthey2006variability}; the State-Trait Anxiety Inventory \citep[STAI,][]{spielberger1983manual} that can be decomposed into two different scores describing the anxious states (STAI-s), reflecting characteristics that can vary with time, and the anxiety traits (STAI-t), reflecting more stable characteristics; and the Pittsburgh Sleep Quality Index \citep[PSQI,][]{buysse1989pittsburgh}. The score ranges of these questionnaires are discretized to obtain clinically relevant categories, and are recorded at different time points, as reported in Supplementary Material Table 2. These five processes represent time-varying categorical observations and are modelled jointly via SUMS, to capture significant relationships between them \citep[see also][]{van2018relation}. In our setting, the four mental health indicators (BDI, EPDS, STAI-s, STAI-t) represent the main clinical responses of interest, while the sleep quality indicator (PSQI) is treated as a time-varying categorical covariate. For all processes, we assume missingness at random and impute missing values at the first time of observation from their full conditionals (see Section 2.2 of Supplementary Material).
We are also provided with information regarding socio-demographic and clinical characteristics, as well as scoring obtained from additional questionnaires measuring personality traits. In particular, we have individual scores for the Big Five Inventory \citep[BFI, ][]{john1999big} (including the scores for \textit{Extraversion}, \textit{Agreeableness}, \textit{Conscientiousness}, \textit{Neuroticism}, \textit{Openness}, and \textit{Liking}) and for the Maternal Childhood Adversity \citep[MCA, ][]{bouvette2015maternal}. Many of the remaining covariates are time-homogeneous categorical, while no time-varying continuous covariates are available. The time-homogeneous continuous covariates are centred and scaled so that each column has null mean and unitary standard deviation, thus estimating the corresponding regression coefficients $\bm \beta^{(h)}$ on the same scale across processes. The full set of covariates ($g^{(h)} = 22$, for $h = 1, \dots, 4$, including dummy coding for the categorical ones) is described in more details in Supplementary Material Table 3, and is included in the specification of the four psychometric processes, but not of PSQI.

\paragraph{Full model specification}
We describe the full model used in the application presented in this section, which is the same implemented in the sensitivity analysis on the hyperparameters $\Lambda$ and $\gamma_S$ appearing in Supplementary Section 3. For each $i = 1, \ldots, N$: 
\begin{eqnarray}\label{eq:MixtureModel1}
	&&\bm Y^{(1)}_i, \dots, \bm Y^{(4)}_i, \bm H_i \mid \{\bm \lambda^{(h)}_i, \bm \beta^{(h)}, h = 1, \ldots, 4\}, \bm \lambda^H_i \ind \prod_{h = 1}^{4} p(\bm Y^{(h)}_i \mid \bm \lambda^{(h)}_i, \bm \beta^{(h)})  p(\bm H_i \mid \bm \lambda^H_i) \nonumber \\
	&&\log \left(\lambda^{(h)}_i(r,s) \right) = \phi^{\star,h}_{c_i}(r,s)  + \bm X^{(h)}_i \bm \beta^{(h)}_{rs}, \ r \rightarrow s, \quad h = 1, \dots, 4  \nonumber\\
	&&\log \left(\lambda^H_i(r,s) \right) = \phi^{\star,H}_{c_i}(r,s), \ r \rightarrow s  \nonumber\\
	&& \bm \beta^{(h)} \sim \text{MN}_{g^{(h)} \times d^{(h)}(d^{(h)}-1)}(\bm 0, \bm U_{\bm \beta^{(h)}}, \bm V_{\bm \beta^{(h)}}), \quad h = 1, \dots, 4 \nonumber \\
	&&\bm \phi^\star_m = (\bm \phi^{\star,1}_m, \dots, \bm \phi^{\star,4}_m, \bm \phi^{\star,H}_m) \mid M, \bm \mu, \bm \Omega_G \sim P_0 =\text{N}_{D_p}(\bm \mu, \bm \Omega_G), \quad m = 1, \dots, M \nonumber \\
	&&\bm \mu, \bm \Omega_G \mid G, \bm m_{\mu}, k_0 \sim \text{N}_{D_p}(\bm \mu \mid \bm m_{\bm \mu}, k_0 \Omega_G) G\text{-Wishart}_G(\bm \Omega_G \mid \nu, \bm \Psi)  \\
	&&k_0 \sim \text{Gamma}(a_{k_0}, b_{k_0}) \nonumber \\
	&&\mathbb{P}(c_i = m) \propto S_m, \quad m = 1, \dots, M \nonumber \\
	&&S_1, \dots, S_M \mid M, \gamma_S \iid \text{Gamma}(\gamma_S, 1) \nonumber \\
	&&M - 1 \mid \Lambda \sim \text{Poi}(\Lambda) \nonumber \\
	&&G = f(G_0), \quad p(G_0) \propto \eta^{|E_0|} (1 - \eta)^{\binom{p}{2} - |E_0|} \nonumber
\end{eqnarray}
where we indicate with $\bm \phi_m^\star$ the vectors of unique $\log$-baseline transition rates for the $m$-th component in the model and $M$ is the unknown number of components in the mixture. Here $\bm{c}= \{c_i, i=\ldots, N\}$ represents the allocation vector, i.e. it specifies to which component the $i$-th observation is assigned to, characterised by $\bm \phi_i =\bm \phi_{c_i}^\star$. The probability of $c_i$ being equal to the $m$-th component of the mixture is proportional to the unnormalized weights $S_m$, for $m = 1, \dots, M$. Therefore, due to the discrete property of the mixing measure, the parameters $\bm \phi_i$ are assigned to $K_N$ different clusters, with $K_N \leq M$. We impose a conditionally conjugate hyper-prior on $k_0$, and fix the hyperparameters $\gamma_S, \Lambda$. We refer to \cite{argiento2019infinity} for a thorough discussion on prior specification in mixture models with unknown number of components. However, we point out that the mixture component of the model is specified conditionally to the graph structure $G$. Finally, $\text{MN}_{n \times p}(\bm 0, \bm U, \bm V)$ is the matrix-variate Normal distribution of dimension $n \times p$ centred on the null matrix $\bm 0$ and with covariance matrices $\bm U$ and $\bm V$ of dimensions $n \times n$ and $p \times p$, respectively.

\paragraph{Hyper-Prior elicitation}\label{SMsec:hyper-prior-elicitation}
We need to specify the hyperparameters for the priors in the three components of the model:  the transition rates, the  mixture model with random number of components and the graphical model. 

In order to induce sparsity in the graph structure and identify meaningful relationship between the SUMS processes, we set the a-priori probability of edge inclusion to $\eta = 0.1$. The hyperparameters of the centring measure $P_0$ are such that $\bm m_{\bm \mu} = \bm 0$, $k_0 \sim \text{Gamma}(1,1)$, $\nu = D_p + 2$ and $\bm \Psi = \mathbb{I}_{D_p} / \nu$, where $\mathbb{I}_p$ is the identity matrix of size $p$. In the case of a full graph $G$, the latter corresponds to $\mathbb{E}(\bm \Omega_G\mid G) = \mathbb{I}_{D_p}$. The regression coefficients $\bm \beta^{(h)}$ are a-priori independent and identically distributed, i.e. $\bm U_{\bm \beta^{(h)}} = \bm V_{\bm \beta^{(h)}} = \mathbb{I}_{g^{(h)} d^{(h)} (d^{(h)} - 1)}$, for $h = 1, \ldots,4$. The mixture prior for the $\log$-baseline transition rates $\bm \phi^\star_1, \dots, \bm \phi^\star_M$ is controlled by the hyperparameters $\Lambda$ and $\gamma_S$. These parameters determine the distribution of the number of components and the corresponding allocation of the subjects, and are the object of an extensive sensitivity analysis presented in Supplementary Section 3. In this application, we fix these parameters to $\Lambda = 0.01$ and $\gamma_S = 0.1$.

\paragraph{Posterior inference}
We run the MCMC algorithm described in Section \ref{sec:mcmc-algorithm} for 50000 iterations, after an initial burn-in period of 1000 iterations used to initialise the adaptive Metropolis-Hastings, discarding 40000 iterations as burn-in and thinning every 2, obtaining a final sample of 5000.

\begin{figure}[ht]
	\centering
	\includegraphics[width=0.65\textwidth]{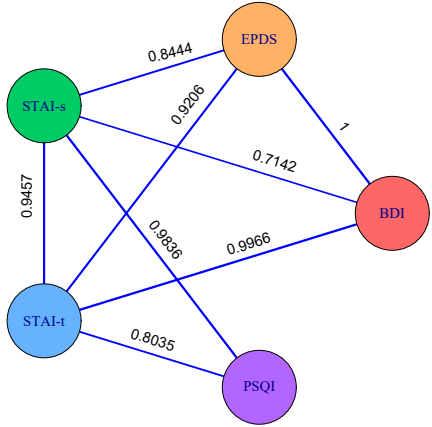}
    \caption{Posterior median graph of $G_0$: each edge included in the graph has posterior edge inclusion probability greater than 0.5. Each edge of the median graph is labelled with the corresponding posterior edge inclusion probability.}
	\label{fig:Graph_G0}
\end{figure}

We explore the relationship between the multi-state processes by imposing dependencies via the graphical model approach described in Section \ref{sec:GGM}. Inference on the posterior distribution of the graphical structure $G_0$ is obtained by reporting the posterior edge inclusion probability for each pair of nodes. In Figure \ref{fig:Graph_G0} we report the posterior median graph, obtained by including only those edges with posterior edge inclusion probability greater than 0.5 \citep{barbieri2004optimal}. The four clinical mental health indicators BDI, EPDS, STAI-s and STAI-t are strongly associated, presenting a clique in the posterior median graph. Interestingly, the sleep quality index PSQI is only related to the anxiety indices STAI-s and STAI-t, forming a clique as well.  Links between probable anxiety and sleeping quality have been reported in previous studies \citep{swanson2011relationships, ibrahim2012sleep}, and it is confirmed by our findings.  Moreover, as previously reported, poor sleep quality may feed into poor emotional and mental health states \citep{ruiz2015sleep, osnes2019insomnia}.

\begin{figure}[ht]
	\centering
	\includegraphics[width=1\textwidth]{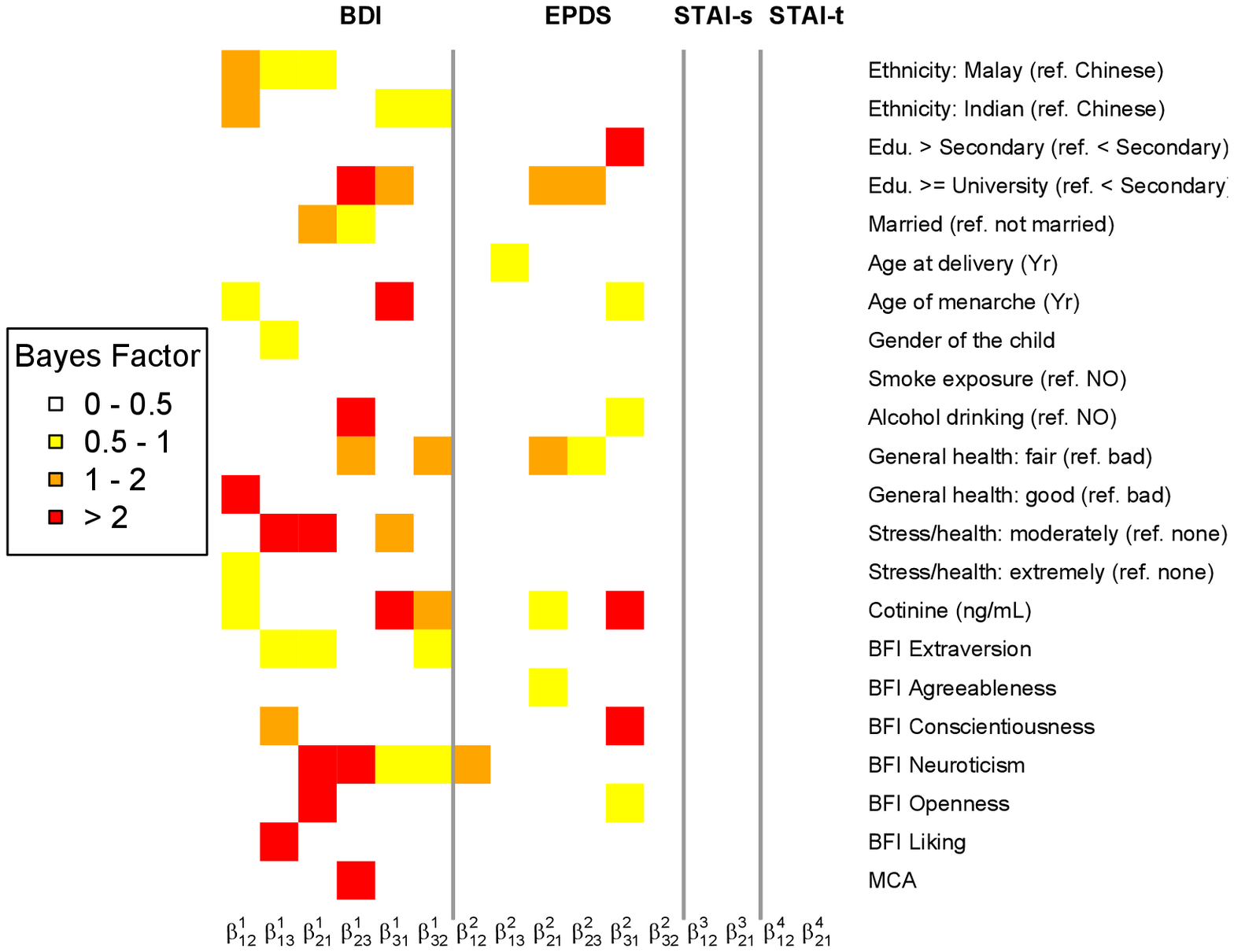}
	\caption{Heatmap of Bayes Factors ($- \log_{10}(\text{BF}^{(h)}_{jk})$) for the individual regression coefficients $\beta^{(h)}_{jk}$, for $j = 1, \dots, g^{(h)}$, $k = 1, \dots, d^{(h)}(d^{(h)} - 1)$ and $h = 1, \dots, 4$. Each row refers to a different covariate included in the model. Each column is associated with a possible transition for each process, excluding PSQI which is modelled as an explanatory factor.}
	\label{fig:Beta_BFs}
\end{figure}

Another important aspect of the proposed model is the possibility of including covariates in the specification of the transition rates via \eqref{eq:Lambdas_Y}. Posterior inference on the coefficient $\bm \beta^{(h)}$, for $h = 1, \dots, 4$ is not trivial, due to the high number of parameters involved. The importance of each covariate can be assessed through Bayes Factors (BF), defined as the ratio of the marginal contributions derived  from  the model with the corresponding regression coefficient set to zero versus the full model \citep{kass1995bayes}. Closed form expressions for the Bayes Factor under the SUMS model are not available, and thus we use the Savage-Dickey density ratio method  \citep{wagenmakers2010bayesian,verdinelli1995computing}. The applicability of this method is guaranteed by the component-wise assumption of independence a-priori for the regression coefficients $\bm \beta^{(h)}$, for $h = 1, \dots, 4$ (see the full model specification in \eqref{eq:MixtureModel1}). For each process $h$, the values of $-\log_{10}\left(\text{BF}^{(h)}_{jk}\right)$ are reported in the heatmap of Figure \ref{fig:Beta_BFs}, for $j = 1, \dots, g^{(h)}$ and $k = 1, \dots, d^{(h)} (d^{(h)} - 1)$. The magnitude of $-\log_{10}\left(\text{BF}^{(h)}_{jk}\right)$ measures the evidence in favour of the full model \citep{kass1995bayes}. The majority of the coefficients is characterized by a low value of $-\log_{10}\left(\text{BF}^{(h)}_{jk}\right)$, supporting the hypothesis of no association, particularly in the case of the STAI processes. However, some coefficients are characterized by $-\log_{10}(BF)$ values above 1 or 2, indicating strong evidence in support of the inclusion of the corresponding covariate in the specific process. Of particular interest are the coefficients relative to the BFI and MCA scores, representing different traits of personality, trauma and parental relationship. We present the posterior mean and 95\% credible intervals of the regression coefficients relative to BFI and MCA in details in Figures 5 of Supplementary Material Section 6. 

The personality traits of the mothers as described by the BFI scores have been previously associated with increased likelihood for both antenatal and postnatal mood disorder traits \citep{ritter2000stress, leigh2008risk}. Our analysis supports this as BFI traits have a relevant impact on both BDI and EPDS (95\% credible interval does not contain zero). An interesting result appears through the estimates of the BFI's \textit{Neuroticism} dimension, which characterizes transitions 2 $\rightarrow$ 3 (deterioration, positive regression coefficient) and 3 $\rightarrow$ 1 (improvement, negative regression coefficient) in both BDI and EPDS scores, indicating that higher \textit{Neuroticism} scores are associated with higher depressive symptoms during the peripartum period \citep{kitamura1993psychological, o1996rates}. On the other hand,  \textit{Openness} and \textit{Conscientiousness} in EPDS (see Supplementary Figure 5)  positively influence the transition 3 $\rightarrow$ 1 (improvement). We also notice the effects of BFI's \textit{Extraversion} and \textit{Agreeableness} differ for BDI and EPDS's transitions. This could be explained by the fact that the social behaviors associated with \textit{Extraversion} and \textit{Agreeableness} are distinct \citep{tobin2000personality, jensen2001agreeableness}. Extraverts tend to actively seek out social interactions, whereas people scoring high on \textit{Agreeableness} prefer harmonious relationships. Maternal history of developmental adversity is linked to increased risk for depression \citep{leigh2008risk}, of which childhood abuse is a strong risk factor \citep{seng2014complex}, as highlighted by the importance of the MCA covariate for the transition 2 $\rightarrow$ 3 (deterioration) in BDI (see Figure \ref{fig:Beta_BFs} and Supplementary Figure 5). This result is also confirmed by  \cite{mandelli2015role} who found that women who were victims of childhood neglect or abuse were at least twice as likely to suffer from depression. The quality of relationship with the women’s parents may also contribute to maternal developmental adversity. Mothers who received low parental care and high control during childhood are at risk for peripartum anxiety \citep{grant2012parental} and depression \citep{mcmahon2005psychological}. 

\begin{figure}[ht]
	\includegraphics[width=0.75\textwidth, angle = -90]{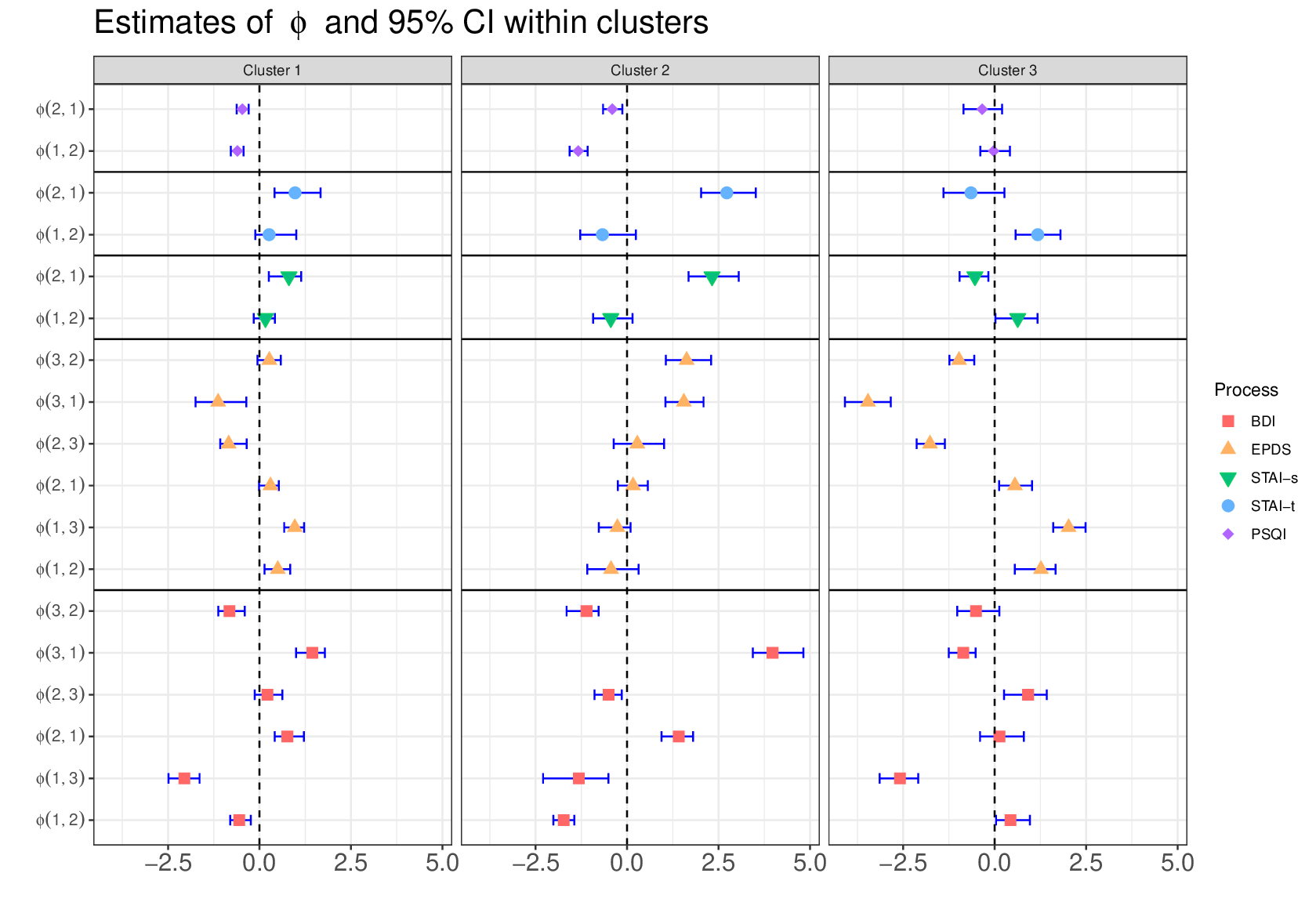}
	\caption{Posterior means and 95\% credible intervals of the instantaneous $\log$-transition rates $\phi^{(h)}(r,s)$ for each process $h = 1, \dots, p_0$. The vertical dashed lines represent the value 0, while the horizontal continuous lines divide the estimates for the five processes. The estimates are obtained by fixing the partition of the subjects to the Binder's partition, and re-running the algorithm for the conditional model. Each sub-plot refers to one of the clusters in the fixed partition.}
	\label{fig:phi_inClusters}
\end{figure}

The choice of the mixture prior \eqref{eq:NIFPP_model} as random effect distribution for the vector of $\log$-transition rates $\bm \phi_1, \dots, \bm \phi_N$ allows for clustering of the subjects. Inference on the random partition is shown in Supplementary Material Figure 6, where the posterior distributions of the number of clusters, components and of the co-clustering probabilities are reported. An estimate of the random partition induced on the subjects under study is obtained by minimizing the Binder's loss function \citep{Binder_1978} with equal costs. We obtain a partition with three clusters, which also corresponds to the posterior mode of the number of clusters. The three clusters contain 140, 126 and 35 subjects, respectively, and are labelled according to their sizes in decreasing order. In Figure \ref{fig:phi_inClusters} we report the posterior distribution of $\phi^{(h)}(r,s)$ conditional on the Binder's partition, for $r \rightarrow s$ and $h = 1, \dots, p$. Cluster-specific estimates of transition rates differ among clusters (see Supplementary Material Section 7 for a discussion). For instance, transition rates corresponding to improvement in the BDI or EPDS scores are higher in Clusters 1 and 2 rather than Cluster 3. The binary processes (STAI-s, STAI-t and PSQI) also seem to present differences between clusters in the same direction, identifying Cluster 3 as the one most prone to a deterioration of the mental health status of its subjects.

\section{Conclusions}\label{sec:Conclusions}
Observations on time-evolving related processes are very common in  biomedical applications and beyond. In this work, we present a Bayesian semiparametric approach for joint modelling of several multi-state Markov processes describing an individual's transitions between different states in continuous time. The proposed model builds on the multi-state Markov models, GGM and PPMx literature.  The different multi-state processes are linked by imposing a flexible prior distribution for the instantaneous transition rates, which allows for  data-driven clustering of the subjects. The dependence among the processes is  captured  by a graph and posterior inference is performed through a  tailored MCMC algorithm.

The proposed model finds wide applicability, due to its flexibility, interpretability and relative ease of computations. 
In this work, we analyse data  from the GUSTO cohort study with the aim of understanding the evolution and relationships between mental health indicators over time. Our findings are in agreement with existing medical literature and shed more light on the influence of childhood and parental factors on mental health progression.  Potential extensions  include  higher order Markov dependency and joint modelling of multi-state  processes and continuous longitudinal trajectories.

A possible alternative to our approach is to represent the categorical covariates with continuous Gaussian latent variables linked to the categorical outcome by thresholding  \citep{albert1993bayesian}, allowing for the inclusion of a time component through auto-regressive terms in the likelihood \citep[e.g.][]{ barcella2018modelling}. To the best of our knowledge, this strategy has not been employed in the context of multi-state models, and it represents an interesting direction for future developments. However, this formulation could suffer from limited interpretability \citep{garcia2007conditional} and could induce further computational challenges \citep{zhang2006sampling}.

\bibliographystyle{plainnat}
\bibliography{Biblio}

\begin{thebibliography}{60}
\providecommand{\natexlab}[1]{#1}
\providecommand{\url}[1]{\texttt{#1}}
\expandafter\ifx\csname urlstyle\endcsname\relax
  \providecommand{\doi}[1]{doi: #1}\else
  \providecommand{\doi}{doi: \begingroup \urlstyle{rm}\Url}\fi

\bibitem[Albert and Chib(1993)]{albert1993bayesian}
James~H Albert and Siddhartha Chib.
\newblock Bayesian analysis of binary and polychotomous response data.
\newblock \emph{Journal of the American statistical Association}, 88\penalty0
  (422):\penalty0 669--679, 1993.

\bibitem[Andersen et~al.(2012)Andersen, Borgan, Gill, and
  Keiding]{andersen2012statistical}
Per~K Andersen, Ornulf Borgan, Richard~D Gill, and Niels Keiding.
\newblock \emph{Statistical models based on counting processes}.
\newblock Springer Science \& Business Media, 2012.

\bibitem[Argiento and De~Iorio(2019)]{argiento2019infinity}
R.~Argiento and M.~De~Iorio.
\newblock Is infinity that far? a bayesian nonparametric perspective of finite
  mixture models.
\newblock \emph{arXiv preprint arXiv:1904.09733}, 2019.

\bibitem[Armstrong et~al.(2009)Armstrong, Carter, Wong, and
  Kohn]{armstrong2009}
H.~Armstrong, C.~K Carter, K.~F.~K. Wong, and R.~Kohn.
\newblock Bayesian covariance matrix estimation using a mixture of decomposable
  graphical models.
\newblock \emph{Statistics and Computing}, 19\penalty0 (3):\penalty0 303--316,
  2009.

\bibitem[Atchad{\'e} et~al.(2005)Atchad{\'e}, Rosenthal,
  et~al.]{atchade2005adaptive}
Yves~F Atchad{\'e}, Jeffrey~S Rosenthal, et~al.
\newblock On adaptive markov chain monte carlo algorithms.
\newblock \emph{Bernoulli}, 11\penalty0 (5):\penalty0 815--828, 2005.

\bibitem[Banterle et~al.(2018)Banterle, Bottolo, Richardson, Ala-Korpela,
  J{\"a}rvelin, and Lewin]{banterle2018sparse}
Marco Banterle, Leonardo Bottolo, Sylvia Richardson, Mika Ala-Korpela,
  MR~J{\"a}rvelin, and Alex Lewin.
\newblock Sparse variable and covariance selection for high-dimensional
  seemingly unrelated bayesian regression.
\newblock \emph{bioRxiv}, page 467019, 2018.

\bibitem[Barbieri et~al.(2004)Barbieri, Berger, et~al.]{barbieri2004optimal}
Maria~Maddalena Barbieri, James~O Berger, et~al.
\newblock Optimal predictive model selection.
\newblock \emph{The annals of statistics}, 32\penalty0 (3):\penalty0 870--897,
  2004.

\bibitem[Barcella et~al.(2017)Barcella, De~Iorio, and
  Baio]{barcella2017comparative}
William Barcella, Maria De~Iorio, and Gianluca Baio.
\newblock A comparative review of variable selection techniques for covariate
  dependent dirichlet process mixture models.
\newblock \emph{Canadian Journal of Statistics}, 45\penalty0 (3):\penalty0
  254--273, 2017.

\bibitem[Barcella et~al.(2018)Barcella, Iorio, and
  Malone-Lee]{barcella2018modelling}
William Barcella, Maria~De Iorio, and James Malone-Lee.
\newblock Modelling correlated binary variables: an application to lower
  urinary tract symptoms.
\newblock \emph{Journal of the Royal Statistical Society: Series C (Applied
  Statistics)}, 67\penalty0 (4):\penalty0 1083--1100, 2018.

\bibitem[Beck et~al.(1961)Beck, Ward, Mendelson, Mock, and
  Erbaugh]{Beck_etal_1961}
A.~T. Beck, C.~H. Ward, M.~Mendelson, J.~Mock, and J.~Erbaugh.
\newblock {An Inventory for Measuring Depression}.
\newblock \emph{Archives of General Psychiatry}, 4\penalty0 (6):\penalty0
  561--571, 06 1961.
\newblock ISSN 0003-990X.
\newblock \doi{10.1001/archpsyc.1961.01710120031004}.
\newblock URL \url{https://doi.org/10.1001/archpsyc.1961.01710120031004}.

\bibitem[Binder(1978)]{Binder_1978}
David~A Binder.
\newblock Bayesian cluster analysis.
\newblock \emph{Biometrika}, 65\penalty0 (1):\penalty0 31--38, 1978.

\bibitem[Bouvette-Turcot et~al.(2015)Bouvette-Turcot, Fleming, Wazana,
  Sokolowski, Gaudreau, Gonzalez, Deslauriers, Kennedy, Steiner, Meaney,
  et~al.]{bouvette2015maternal}
Andr{\'e}e-Anne Bouvette-Turcot, AS~Fleming, A~Wazana, MB~Sokolowski,
  H~Gaudreau, A~Gonzalez, J~Deslauriers, JL~Kennedy, M~Steiner, MJ~Meaney,
  et~al.
\newblock Maternal childhood adversity and child temperament: An association
  moderated by child 5-httlpr genotype.
\newblock \emph{Genes, Brain and Behavior}, 14\penalty0 (3):\penalty0 229--237,
  2015.

\bibitem[Buysse et~al.(1989)Buysse, Reynolds, Monk, Berman, Kupfer,
  et~al.]{buysse1989pittsburgh}
Daniel~J Buysse, Charles~F Reynolds, Timothy~H Monk, Susan~R Berman, David~J
  Kupfer, et~al.
\newblock The pittsburgh sleep quality index: a new instrument for psychiatric
  practice and research.
\newblock \emph{Psychiatry res}, 28\penalty0 (2):\penalty0 193--213, 1989.

\bibitem[Cook(1999)]{Cook_1999}
Richard~J Cook.
\newblock A mixed model for two-state markov processes under panel observation.
\newblock \emph{Biometrics}, 55\penalty0 (3):\penalty0 915--920, 1999.

\bibitem[Cox and Miller(1977)]{cox1977theory}
David~Roxbee Cox and Hilton~David Miller.
\newblock \emph{The theory of stochastic processes}, volume 134.
\newblock CRC press, 1977.

\bibitem[De~Iorio et~al.(2018)De~Iorio, Gallot, Valcarcel, and
  Wedderburn]{DeIorio_etal_2018}
Maria De~Iorio, Natacha Gallot, Beatriz Valcarcel, and Lucy Wedderburn.
\newblock A bayesian semiparametric markov regression model for juvenile
  dermatomyositis.
\newblock \emph{Statistics in medicine}, 37\penalty0 (10):\penalty0 1711--1731,
  2018.

\bibitem[Dempster(1972)]{Dempster}
A.~Dempster.
\newblock Covariance selection.
\newblock \emph{Biometrics}, 28:\penalty0 157--175, 1972.

\bibitem[Ferrer et~al.(2016)Ferrer, Rondeau, Dignam, Pickles, Jacqmin-Gadda,
  and Proust-Lima]{ferrer2016joint}
Lo{\"\i}c Ferrer, Virginie Rondeau, James Dignam, Tom Pickles, H{\'e}l{\`e}ne
  Jacqmin-Gadda, and C{\'e}cile Proust-Lima.
\newblock Joint modelling of longitudinal and multi-state processes:
  application to clinical progressions in prostate cancer.
\newblock \emph{Statistics in medicine}, 35\penalty0 (22):\penalty0 3933--3948,
  2016.

\bibitem[Garc{\'\i}a-Zattera et~al.(2007)Garc{\'\i}a-Zattera, Jara, Lesaffre,
  and Declerck]{garcia2007conditional}
Mar{\'\i}a~Jos{\'e} Garc{\'\i}a-Zattera, Alejandro Jara, Emmanuel Lesaffre, and
  Dominique Declerck.
\newblock Conditional independence of multivariate binary data with an
  application in caries research.
\newblock \emph{Computational statistics \& data analysis}, 51\penalty0
  (6):\penalty0 3223--3234, 2007.

\bibitem[Grant et~al.(2012)Grant, Bautovich, McMahon, Reilly, Leader, and
  Austin]{grant2012parental}
Kerry-Ann Grant, Alison Bautovich, Catherine McMahon, Nicole Reilly, Leo
  Leader, and Marie-Paule Austin.
\newblock Parental care and control during childhood: associations with
  maternal perinatal mood disturbance and parenting stress.
\newblock \emph{Archives of women's mental health}, 15\penalty0 (4):\penalty0
  297--305, 2012.

\bibitem[Grimmet and Sterzaker(2001)]{grimmet1992probability}
Geoffrey Grimmet and David Sterzaker.
\newblock \emph{Probability and Random Processes.}
\newblock Oxford University press, 2001.

\bibitem[Haario et~al.(2001)Haario, Saksman, Tamminen,
  et~al.]{haario2001adaptive}
Heikki Haario, Eero Saksman, Johanna Tamminen, et~al.
\newblock An adaptive metropolis algorithm.
\newblock \emph{Bernoulli}, 7\penalty0 (2):\penalty0 223--242, 2001.

\bibitem[Houle et~al.(2010)Houle, Govindaraju, and Omholt]{houle2010phenomics}
David Houle, Diddahally~R Govindaraju, and Stig Omholt.
\newblock Phenomics: the next challenge.
\newblock \emph{Nature reviews genetics}, 11\penalty0 (12):\penalty0 855--866,
  2010.

\bibitem[Ibrahim and Foldvary-Schaefer(2012)]{ibrahim2012sleep}
Sally Ibrahim and Nancy Foldvary-Schaefer.
\newblock Sleep disorders in pregnancy: implications, evaluation, and
  treatment.
\newblock \emph{Neurologic clinics}, 30\penalty0 (3):\penalty0 925--936, 2012.

\bibitem[Jackson et~al.(2011)]{jackson_2011}
Christopher~H Jackson et~al.
\newblock Multi-state models for panel data: the msm package for r.
\newblock \emph{Journal of statistical software}, 38\penalty0 (8):\penalty0
  1--29, 2011.

\bibitem[James et~al.(2009)James, Lijoi, and Pr{\"u}nster]{jamesetal09}
L.~F. James, A.~Lijoi, and I.~Pr{\"u}nster.
\newblock Posterior analysis for normalized random measures with independent
  increments.
\newblock \emph{Scandinavian Journal of Statistics}, 36\penalty0 (1):\penalty0
  76--97, 2009.

\bibitem[Jensen-Campbell and Graziano(2001)]{jensen2001agreeableness}
Lauri~A Jensen-Campbell and William~G Graziano.
\newblock Agreeableness as a moderator of interpersonal conflict.
\newblock \emph{Journal of personality}, 69\penalty0 (2):\penalty0 323--362,
  2001.

\bibitem[John et~al.(1999)John, Srivastava, et~al.]{john1999big}
Oliver~P John, Sanjay Srivastava, et~al.
\newblock The big five trait taxonomy: History, measurement, and theoretical
  perspectives.
\newblock \emph{Handbook of personality: Theory and research}, 2\penalty0
  (1999):\penalty0 102--138, 1999.

\bibitem[Kass and Raftery(1995)]{kass1995bayes}
Robert~E Kass and Adrian~E Raftery.
\newblock Bayes factors.
\newblock \emph{Journal of the american statistical association}, 90\penalty0
  (430):\penalty0 773--795, 1995.

\bibitem[Kim et~al.(2012)Kim, James, and Weissbach]{kim2012bayesian}
Yongdai Kim, Lancelot James, and Rafael Weissbach.
\newblock Bayesian analysis of multistate event history data: beta-dirichlet
  process prior.
\newblock \emph{Biometrika}, 99\penalty0 (1):\penalty0 127--140, 2012.

\bibitem[Kitamura et~al.(1993)Kitamura, Shima, Sugawara, and
  Toda]{kitamura1993psychological}
T~Kitamura, S~Shima, M~Sugawara, and MA~Toda.
\newblock Psychological and social correlates of the onset of affective
  disorders among pregnant women.
\newblock \emph{Psychological medicine}, 23\penalty0 (4):\penalty0 967--975,
  1993.

\bibitem[Kneib and Hennerfeind(2008)]{kneib2008bayesian}
Thomas Kneib and Andrea Hennerfeind.
\newblock Bayesian semi parametric multi-state models.
\newblock \emph{Statistical Modelling}, 8\penalty0 (2):\penalty0 169--198,
  2008.

\bibitem[Leigh and Milgrom(2008)]{leigh2008risk}
Bronwyn Leigh and Jeannette Milgrom.
\newblock Risk factors for antenatal depression, postnatal depression and
  parenting stress.
\newblock \emph{BMC psychiatry}, 8\penalty0 (1):\penalty0 24, 2008.

\bibitem[Mandelli et~al.(2015)Mandelli, Petrelli, and
  Serretti]{mandelli2015role}
L~Mandelli, C~Petrelli, and A~Serretti.
\newblock The role of specific early trauma in adult depression: a
  meta-analysis of published literature. childhood trauma and adult depression.
\newblock \emph{European psychiatry}, 30\penalty0 (6):\penalty0 665--680, 2015.

\bibitem[Matthey et~al.(2006)Matthey, Henshaw, Elliott, and
  Barnett]{matthey2006variability}
S~Matthey, C~Henshaw, S~Elliott, and B~Barnett.
\newblock Variability in use of cut-off scores and formats on the edinburgh
  postnatal depression scale--implications for clinical and research practice.
\newblock \emph{Archives of women's mental health}, 9\penalty0 (6):\penalty0
  309--315, 2006.

\bibitem[McMahon et~al.(2005)McMahon, Barnett, Kowalenko, and
  Tennant]{mcmahon2005psychological}
Catherine McMahon, Bryanne Barnett, Nicholas Kowalenko, and Christopher
  Tennant.
\newblock Psychological factors associated with persistent postnatal
  depression: past and current relationships, defence styles and the mediating
  role of insecure attachment style.
\newblock \emph{Journal of affective disorders}, 84\penalty0 (1):\penalty0
  15--24, 2005.

\bibitem[Meaney et~al.(2018)Meaney, van Lee, Cai, Loy, Tham, Yap, Godfrey,
  Gluckman, Shek, Teoh, Goh, et~al.]{van2018relation}
Michael Meaney, Linde van Lee, Shirong Cai, See~Ling Loy, Elaine~KH Tham,
  Fabian~KP Yap, Keith~M Godfrey, Peter~D Gluckman, Lynette~PC Shek, Oon~Hoe
  Teoh, Daniel~YT Goh, et~al.
\newblock Relation of plasma tryptophan concentrations during pregnancy to
  maternal sleep and mental well-being: The gusto cohort.
\newblock \emph{Journal of affective disorders}, 225:\penalty0 523--529, 2018.

\bibitem[Mohammadi et~al.(2015)Mohammadi, Wit, et~al.]{mohammadi2015bayesian}
Abdolreza Mohammadi, Ernst~C Wit, et~al.
\newblock Bayesian structure learning in sparse gaussian graphical models.
\newblock \emph{Bayesian Analysis}, 10\penalty0 (1):\penalty0 109--138, 2015.

\bibitem[Moler and Van~Loan(2003)]{moler2003nineteen}
Cleve Moler and Charles Van~Loan.
\newblock Nineteen dubious ways to compute the exponential of a matrix,
  twenty-five years later.
\newblock \emph{SIAM review}, 45\penalty0 (1):\penalty0 3--49, 2003.

\bibitem[M{\"u}ller et~al.(1996)M{\"u}ller, Erkanli, and
  West]{muller1996bayesian}
Peter M{\"u}ller, Alaattin Erkanli, and Mike West.
\newblock Bayesian curve fitting using multivariate normal mixtures.
\newblock \emph{Biometrika}, 83\penalty0 (1):\penalty0 67--79, 1996.

\bibitem[M{\"u}ller et~al.(2011)M{\"u}ller, Quintana, and
  Rosner]{muller2011product}
Peter M{\"u}ller, Fernando Quintana, and Gary~L Rosner.
\newblock A product partition model with regression on covariates.
\newblock \emph{Journal of Computational and Graphical Statistics}, 20\penalty0
  (1):\penalty0 260--278, 2011.

\bibitem[O'hara and Swain(1996)]{o1996rates}
Michael~W O'hara and Annette~M Swain.
\newblock Rates and risk of postpartum depression—a meta-analysis.
\newblock \emph{International review of psychiatry}, 8\penalty0 (1):\penalty0
  37--54, 1996.

\bibitem[Osnes et~al.(2019)Osnes, Roaldset, Follestad, and
  Eberhard-Gran]{osnes2019insomnia}
Rannveig~S Osnes, John~Olav Roaldset, Turid Follestad, and Malin Eberhard-Gran.
\newblock Insomnia late in pregnancy is associated with perinatal anxiety: a
  longitudinal cohort study.
\newblock \emph{Journal of affective disorders}, 248:\penalty0 155--165, 2019.

\bibitem[Ritter et~al.(2000)Ritter, Hobfoll, Lavin, Cameron, and
  Hulsizer]{ritter2000stress}
Christian Ritter, Stevan~E Hobfoll, Justin Lavin, Rebecca~P Cameron, and
  Michael~R Hulsizer.
\newblock Stress, psychosocial resources, and depressive symptomatology during
  pregnancy in low-income, inner-city women.
\newblock \emph{Health Psychology}, 19\penalty0 (6):\penalty0 576, 2000.

\bibitem[Ross et~al.(1996)Ross, Kelly, Sullivan, Perry, Mercer, Davis,
  Washburn, Sager, Boyce, and Bristow]{Ross_1996}
Sheldon~M Ross, John~J Kelly, Roger~J Sullivan, William~James Perry, Donald
  Mercer, Ruth~M Davis, Thomas~Dell Washburn, Earl~V Sager, Joseph~B Boyce, and
  Vincent~L Bristow.
\newblock \emph{Stochastic processes}, volume~2.
\newblock Wiley New York, 1996.

\bibitem[Roverato(2002)]{Roverato2002}
A.~Roverato.
\newblock Hyper inverse {W}ishart distribution for non-decomposable graphs and
  its application to {B}ayesian inference for {G}aussian graphical models.
\newblock \emph{Scandinavian Journal of Statistics}, 29\penalty0 (3):\penalty0
  391--411, 2002.

\bibitem[Ruiz-Robledillo et~al.(2015)Ruiz-Robledillo, Can{\'a}rio, Dias,
  Moya-Albiol, and Figueiredo]{ruiz2015sleep}
N~Ruiz-Robledillo, Catarina Can{\'a}rio, CC~Dias, L~Moya-Albiol, and
  B{\'a}rbara Figueiredo.
\newblock Sleep during the third trimester of pregnancy: the role of depression
  and anxiety.
\newblock \emph{Psychology, health \& medicine}, 20\penalty0 (8):\penalty0
  927--932, 2015.

\bibitem[Seng et~al.(2014)Seng, D'Andrea, and Ford]{seng2014complex}
Julia~S Seng, Wendy D'Andrea, and Julian~D Ford.
\newblock Complex mental health sequelae of psychological trauma among women in
  prenatal care.
\newblock \emph{Psychological Trauma: Theory, Research, Practice, and Policy},
  6\penalty0 (1):\penalty0 41, 2014.

\bibitem[Soh et~al.(2014)Soh, Tint, Gluckman, Godfrey, Rifkin-Graboi, Chan,
  St{\"u}nkel, Holbrook, Kwek, Chong, et~al.]{soh2014cohort}
Shu-E Soh, Mya~Thway Tint, Peter~D Gluckman, Keith~M Godfrey, Anne
  Rifkin-Graboi, Yiong~Huak Chan, Walter St{\"u}nkel, Joanna~D Holbrook,
  Kenneth Kwek, Yap-Seng Chong, et~al.
\newblock Cohort profile: Growing up in singapore towards healthy outcomes
  (gusto) birth cohort study.
\newblock \emph{International journal of epidemiology}, 43\penalty0
  (5):\penalty0 1401--1409, 2014.

\bibitem[Spielberger et~al.(1983)Spielberger, Gorsuch, Lushene, Vagg, and
  Jacobs]{spielberger1983manual}
Charles~D Spielberger, Richard~L Gorsuch, R~Lushene, Peter~R Vagg, and Gerard~A
  Jacobs.
\newblock \emph{Manual for the state-trait anxiety inventory}.
\newblock Consulting Psychologists Press, Palo Alto, CA, 1983.

\bibitem[Swanson et~al.(2011)Swanson, Pickett, Flynn, and
  Armitage]{swanson2011relationships}
Leslie~M Swanson, Scott~M Pickett, Heather Flynn, and Roseanne Armitage.
\newblock Relationships among depression, anxiety, and insomnia symptoms in
  perinatal women seeking mental health treatment.
\newblock \emph{Journal of Women's Health}, 20\penalty0 (4):\penalty0 553--558,
  2011.

\bibitem[Thissen and Steinberg(2009)]{thissen2009item}
David Thissen and Lynne Steinberg.
\newblock Item response theory.
\newblock \emph{The Sage handbook of quantitative methods in psychology}, pages
  148--177, 2009.

\bibitem[Tobin et~al.(2000)Tobin, Graziano, Vanman, and
  Tassinary]{tobin2000personality}
Ren{\'e}e~M Tobin, William~G Graziano, Eric~J Vanman, and Louis~G Tassinary.
\newblock Personality, emotional experience, and efforts to control emotions.
\newblock \emph{Journal of personality and social psychology}, 79\penalty0
  (4):\penalty0 656, 2000.

\bibitem[van~den Hout et~al.(2015)van~den Hout, Fox, and
  Klein~Entink]{van_2015}
Ardo van~den Hout, Jean-Paul Fox, and Rinke~H Klein~Entink.
\newblock Bayesian inference for an illness-death model for stroke with
  cognition as a latent time-dependent risk factor.
\newblock \emph{Statistical methods in medical research}, 24\penalty0
  (6):\penalty0 769--787, 2015.

\bibitem[Verdinelli and Wasserman(1995)]{verdinelli1995computing}
Isabella Verdinelli and Larry Wasserman.
\newblock Computing bayes factors using a generalization of the savage-dickey
  density ratio.
\newblock \emph{Journal of the American Statistical Association}, 90\penalty0
  (430):\penalty0 614--618, 1995.

\bibitem[Wagenmakers et~al.(2010)Wagenmakers, Lodewyckx, Kuriyal, and
  Grasman]{wagenmakers2010bayesian}
Eric-Jan Wagenmakers, Tom Lodewyckx, Himanshu Kuriyal, and Raoul Grasman.
\newblock Bayesian hypothesis testing for psychologists: A tutorial on the
  savage--dickey method.
\newblock \emph{Cognitive psychology}, 60\penalty0 (3):\penalty0 158--189,
  2010.

\bibitem[Wang(2010)]{wang2010sparse}
Hao Wang.
\newblock Sparse seemingly unrelated regression modelling: Applications in
  finance and econometrics.
\newblock \emph{Computational Statistics \& Data Analysis}, 54\penalty0
  (11):\penalty0 2866--2877, 2010.

\bibitem[Wang et~al.(2012)Wang, Li, et~al.]{wang2012efficient}
Hao Wang, Sophia~Zhengzi Li, et~al.
\newblock Efficient gaussian graphical model determination under g-wishart
  prior distributions.
\newblock \emph{Electronic Journal of Statistics}, 6:\penalty0 168--198, 2012.

\bibitem[Zellner(1963)]{zellner1963estimators}
Arnold Zellner.
\newblock Estimators for seemingly unrelated regression equations: Some exact
  finite sample results.
\newblock \emph{Journal of the American Statistical Association}, 58\penalty0
  (304):\penalty0 977--992, 1963.

\bibitem[Zhang et~al.(2006)Zhang, Boscardin, and Belin]{zhang2006sampling}
Xiao Zhang, W~John Boscardin, and Thomas~R Belin.
\newblock Sampling correlation matrices in bayesian models with correlated
  latent variables.
\newblock \emph{Journal of Computational and Graphical Statistics}, 15\penalty0
  (4):\penalty0 880--896, 2006.

\end{thebibliography}


\begin{thebibliography}{12}
\providecommand{\natexlab}[1]{#1}
\providecommand{\url}[1]{\texttt{#1}}
\expandafter\ifx\csname urlstyle\endcsname\relax
  \providecommand{\doi}[1]{doi: #1}\else
  \providecommand{\doi}{doi: \begingroup \urlstyle{rm}\Url}\fi

\bibitem[Argiento and De~Iorio(2019)]{argiento2019infinity}
R.~Argiento and M.~De~Iorio.
\newblock Is infinity that far? a bayesian nonparametric perspective of finite
  mixture models.
\newblock \emph{arXiv preprint arXiv:1904.09733}, 2019.

\bibitem[Binder(1978)]{Binder_1978}
David~A Binder.
\newblock Bayesian cluster analysis.
\newblock \emph{Biometrika}, 65\penalty0 (1):\penalty0 31--38, 1978.

\bibitem[Carvalho et~al.(2007)Carvalho, Massam, and West]{Carvalho_etal2007}
Carlos~M Carvalho, H{\'e}l{\`e}ne Massam, and Mike West.
\newblock Simulation of hyper-inverse wishart distributions in graphical
  models.
\newblock \emph{Biometrika}, 94\penalty0 (3):\penalty0 647--659, 2007.

\bibitem[De~Iorio et~al.(2018)De~Iorio, Gallot, Valcarcel, and
  Wedderburn]{DeIorio_etal_2018}
Maria De~Iorio, Natacha Gallot, Beatriz Valcarcel, and Lucy Wedderburn.
\newblock A bayesian semiparametric markov regression model for juvenile
  dermatomyositis.
\newblock \emph{Statistics in medicine}, 37\penalty0 (10):\penalty0 1711--1731,
  2018.

\bibitem[Giudici and Green(1999)]{Giudici_Green1999}
Paolo Giudici and PJ~Green.
\newblock Decomposable graphical gaussian model determination.
\newblock \emph{Biometrika}, 86\penalty0 (4):\penalty0 785--801, 1999.

\bibitem[Haario et~al.(2001)Haario, Saksman, Tamminen,
  et~al.]{haario2001adaptive}
Heikki Haario, Eero Saksman, Johanna Tamminen, et~al.
\newblock An adaptive metropolis algorithm.
\newblock \emph{Bernoulli}, 7\penalty0 (2):\penalty0 223--242, 2001.

\bibitem[Jackson et~al.(2011)]{jackson_2011}
Christopher~H Jackson et~al.
\newblock Multi-state models for panel data: the msm package for r.
\newblock \emph{Journal of statistical software}, 38\penalty0 (8):\penalty0
  1--29, 2011.

\bibitem[Lenkoski(2013)]{Lenkoski_2013}
Alex Lenkoski.
\newblock A direct sampler for g-wishart variates.
\newblock \emph{Stat}, 2\penalty0 (1):\penalty0 119--128, 2013.

\bibitem[Mohammadi et~al.(2015)Mohammadi, Wit, et~al.]{mohammadi2015bayesian}
Abdolreza Mohammadi, Ernst~C Wit, et~al.
\newblock Bayesian structure learning in sparse gaussian graphical models.
\newblock \emph{Bayesian Analysis}, 10\penalty0 (1):\penalty0 109--138, 2015.

\bibitem[M{\"u}ller et~al.(1996)M{\"u}ller, Erkanli, and
  West]{muller1996bayesian}
Peter M{\"u}ller, Alaattin Erkanli, and Mike West.
\newblock Bayesian curve fitting using multivariate normal mixtures.
\newblock \emph{Biometrika}, 83\penalty0 (1):\penalty0 67--79, 1996.

\bibitem[M{\"u}ller et~al.(2011)M{\"u}ller, Quintana, and
  Rosner]{muller2011product}
Peter M{\"u}ller, Fernando Quintana, and Gary~L Rosner.
\newblock A product partition model with regression on covariates.
\newblock \emph{Journal of Computational and Graphical Statistics}, 20\penalty0
  (1):\penalty0 260--278, 2011.

\bibitem[Roverato(2002)]{Roverato2002}
A.~Roverato.
\newblock Hyper inverse {W}ishart distribution for non-decomposable graphs and
  its application to {B}ayesian inference for {G}aussian graphical models.
\newblock \emph{Scandinavian Journal of Statistics}, 29\penalty0 (3):\penalty0
  391--411, 2002.

\end{thebibliography}

%
%
%

\end{document}


\title{Seemingly Unrelated Multi-State processes: a Bayesian semiparametric approach \\ Supplementary Material}
	
	\author[1]{Andrea Cremaschi}
	\author[2,1,3,4]{Raffele Argiento}
	\author[2,1,3,4]{Maria De Iorio}
	\author[1]{Cai Shirong}
	\author[2,1]{Yap Seng Chong}
	\author[1,2,5]{Michael J. Meaney}
	\author[1]{Michelle Z. L. Kee}
	
	\affil[1]{Singapore Institute for Clinical Sciences, A*STAR, Singapore}
	\affil[2]{Yong Loo Lin School of Medicine, National University of Singapore, Singapore}
	\affil[3]{Division of Science, Yale-NUS College, Singapore}
	\affil[4]{Department of Statistical Science, University College London, UK}
	\affil[5]{Department of Psychiatry, McGill University, Montreal, Canada}
	
	\date{}
	
	\maketitle
	
	\textbf{Summary}: This Supplementary Material contains additional information regarding the manuscript titled ``Seemingly Unrelated Multi-State processes: a Bayesian nonparametric approach''. In particular, it is organized as follows: Section 1 contains a short proof of Eq.(5) reported in the main text; Section 2 describes the algorithm devised for posterior inference for the proposed model, including the description on how to update the graphical strictures and the hyperparameters of the model; Section 3 shows a sensitivity analysis on the hyperparameters of the semiparametric mixture model when applied to the GUSTO cohort study; Section 4 presents a simulation study offering a comparison of the proposed model with competitors; while Sections 5 and 6 present details on the GUSTO data and additional figures discussed in the application.

\clearpage
\section{Proof of Eq.(5)}\label{secSM:proof_Eq_5}
As shown in Section 2.4 of the main manuscript, the SUMS model presents an interesting relationship with the PPMx framework proposed by \cite{muller1996bayesian, muller2011product}. We show this by modelling the set of processes $(Y^{(1)}(t), \dots, Y^{(p_Y)}(t))$ as responses and the set $(H^{(1)}(t), \dots, H^{(p_H)}(t))$ as explanatory factors, and by specifying a suitable joint probability model based on the likelihood described in Eq.(2) of the main manuscript. Then, considering the partition $\rho_N$ of the indices $\{1, \dots, N\}$, we have that in the PPMx framework:
\begin{equation}\label{SMeq:rho_H_1}
	p\left(\rho_N \mid \bm H \right) = V(N, K_N) \prod_{j = 1}^{K_N} \mathcal{C}(C_j) \mathcal{G}(\bm H^\star_j)
\end{equation}
where $\mathcal{C}(C_j)$ is the \textit{cohesion}, i.e. a function of the $j$-th cluster $C_j$, $\mathcal{G}(\bm H^\star_j)$ is the \textit{similarity}, i.e. a function of the array of covariates corresponding to the subjects in cluster $j$, denoted as $\bm H^\star_j$, for $j = 1, \dots, K_N$, and $V(N, K_N)$ is a constant depending only on the sample size $N$ and the number of clusters $K_N$. The cohesion function $\mathcal{C}$ expresses prior information about the partition, such as the average size of a cluster, while the similarity function $\mathcal{G}$ captures the contribution of the covariates to the clustering structure. The specification of $\mathcal{G}$ allows subjects with similar covariates to be more likely assigned to the same cluster.

Eq. \eqref{SMeq:rho_H_1} for the proposed model can be obtained by first considering the following joint distribution:
\begin{gather}
	\mathcal{L}\left(\bm Y, \bm H, \bm \lambda^Y, \bm \lambda^H, \rho_N, M \right) \propto \nonumber \\
	p(\rho_N) \prod_{j = 1}^{K_N} \left( \prod_{i \in C_j} \left( p\left(\bm Y_i \mid \bm \lambda^Y_i \right) p\left(\bm H_i \mid \bm \lambda^H_i \right) \right) P_0(\bm \phi^\star_j)  \right) \prod_{m = K_N+1}^M P_0(\bm \phi^\star_m) \label{SMeq:joint_all}
\end{gather}
where $\bm Y$, $\bm H$, $\bm \lambda^Y$ and $\bm \lambda^H$ indicate the multi-dimensional arrays containing the observation vectors $\bm Y^{(h)}_i$ and $\bm H^{(l)}_i$ and the instantaneous transition rate vectors $\bm \lambda^{(h)}_i$ and $\bm \lambda^{(l)}_i$, respectively. Recall that the $i$-th instantaneous transition rates $\bm \lambda^{(h)}_i$ and $\bm \lambda^{(l)}_i$ depend on the baseline $\log$-transition rates $\bm \phi_i = (\bm \phi^{(1)}_i, \dots, \bm \phi^{(p_Y)}_i, \bm \phi^{(p_Y + 1)}_i, \dots, \bm \phi^{(p)}_i)$, for $i = 1, \dots, N$. The vectors $\bm \phi_i$ are modelled using the mixture prior in (3) (main text) with random number of components and with base measure $P_0(\bm \phi^\star_m)$ for the corresponding unique values $\bm \phi^\star_m = (\bm \phi^{\star, (1)}_m, \dots, \bm \phi^{\star, (p_Y)}_m, \bm \phi^{\star, (p_Y + 1)}_m, \dots, \bm \phi^{\star, (p)}_m)$, for $m = 1, \dots, M$. Following \cite{argiento2019infinity}, the law of the partition $\rho_N$ a-priori is:
\begin{equation}\label{SMeq:partition}
	p\left( \rho_N \right) = V(N,K_N) \prod_{j = 1}^{K_N} \frac{\Gamma(\gamma_S + n_j)}{\Gamma(\gamma_S)}
\end{equation}
where $V(N,K_N)$ is a constant given in integral form in \cite{argiento2019infinity}, $n_j$ denotes the number of elements in the $j$-th cluster, for $j = 1, \dots, K_N$, and $\Gamma(x)$ denotes Euler's gamma function of argument $x > 0$. 

In \eqref{SMeq:joint_all}, we can integrate out $\bm Y$ and the components of $\bm \phi^\star_j$ corresponding to the response processes, i.e. $(\bm \phi^{\star,(1)}_j, \dots, \bm \phi^{\star,(p_Y)}_j)$ for $j = 1, \dots, K_N$, as well as the vectors corresponding to the non-allocated process $\bm \phi^\star_m$ for $m = K_N+1, \dots, M$ \citep[see][for details]{argiento2019infinity}. This reduces to the following marginal law:
\begin{gather}
	\mathcal{L}\left(\bm H, \bm \lambda^H, \rho_N \right) \propto \nonumber \\
	V(N,K_N)\prod_{j = 1}^k \frac{\Gamma(\gamma_S + n_j)}{\Gamma(\gamma_S)} \left( \prod_{i \in C_j}  p\left(\bm H_i \mid \bm \lambda^H_i \right) \right) P_0(\bm \phi^{\star, (p_Y + 1)}_j, \dots, \bm \phi^{\star, (p)}_j) \label{SMeq:joint_rho_H}
\end{gather}
where $P_0(\bm \phi^{\star, (p_Y + 1)}_j, \dots, \bm \phi^{\star, (p)}_j)$ indicates the baseline measure for the components of the vector of log baseline transition rates after marginalizing with respect to the ones corresponding to the response processes, i.e. the first $p_Y$ ones. The marginal law \eqref{SMeq:joint_rho_H} easily yields the expression for \eqref{SMeq:rho_H_1} we are interested in, by marginalizing with respect to $(\bm \phi^{\star, (p_Y + 1)}_j, \dots, \bm \phi^{\star, (p)}_j)$:
\begin{gather}
	p\left(\rho_N \mid \bm H \right) \propto p\left( \rho_N \right) p\left(\bm H \mid \rho_N \right) = \nonumber \\
	V(N, K_N) \prod_{j = 1}^{K_N} \frac{\Gamma(\gamma_S + n_j)}{\Gamma(\gamma_S)} \int \left( \prod_{i \in C_j}  p\left(\bm H_i \mid \bm \lambda^H_i \right) \right) P_0(d\bm \phi^{\star, (p_Y + 1)}_j, \dots, d\bm \phi^{\star, (p)}_j) \label{SMeq:rho_H_2}
\end{gather}
Note how the expression of the similarity function $\mathcal{G}(\bm H^\star_j)$ corresponding to the integral in \eqref{SMeq:rho_H_2} is not known in closed form, due to the non-conjugacy of the proposed model. This differs from the original PPMx specification, where the similarity function is usually obtained from a conjugate model for the covariates vector via marginalization to simplify computations. In the proposed approach, the evaluation of \eqref{SMeq:rho_H_2} would require an expensive numerical approximation. For this reason, as detailed in the following Section, we resort to a conditional MCMC algorithm, not requiring the evaluation of the integral in \eqref{SMeq:rho_H_2}.

\section{Computational details}\label{SMsec:FullCond_and_Alg}
We give the steps of MCMC algorithm. First, we discuss the update of the precision matrix $\bm \Omega_G$ and of the graphs $G$ and $G_0$. To this end, we need to extend the birth-and-death algorithm of \cite{mohammadi2015bayesian} to our modelling framework. The main novelty is in the transition kernel defined on graph space, which involves moving entire cliques in $G$ instead of single edges, necessary due to the presence of the bijection between $G_0$ and $G$. Then, we present the updates of the other parameters in the model, including the regression coefficients, the partition, and the number of components in the mixture.

\subsection{Updating $(\Omega_G, G, G_0)$}
Recall that $\bm \phi^{\star} = (\bm \phi^{\star}_1, \dots, \bm \phi^{\star}_M)$ is the vector of unique $\log$-baseline transition rates associated to the $M$ components of the mixture specified in Eq.(1) in the main manuscript. We want to sample from the following full-conditional distribution:
$$
p(\bm \Omega_G, G, G_0 \mid \bm \phi^{\star}, \bm \mu, \kappa_0, \eta, \bm Y) \propto p(\bm \phi^{\star} \mid \bm \Omega_G, \bm \mu, \kappa_0) p(\bm \Omega_G \mid G) p(G \mid G_0, \eta) p(G_0 \mid \eta)
$$
recalling that $p(G \mid G_0, \eta) = \delta_{f(G_0)}(G)$, since the map linking $G_0$ and $G$ is a bijection. It is well known \citep{Roverato2002} that the $G$-Wishart prior for the precision matrix $\bm \Omega_G$ is conjugate to the multivariate normal. The full conditional for $\bm \Omega_G$, after marginalizing with respect to the mean vector $\bm \mu$, is then:
\begin{equation}\label{eq:FC_OmegaG_marg}
p(\bm \Omega_G \mid G, \bm \phi^{\star}) = G\text{-Wishart}(\nu^*, \bm \Psi^*)
\end{equation}
with $\nu^* = \nu + M$ and $\bm \Psi^* = \bm \Psi + \sum_{m = 1}^M (\bm \phi^{\star}_m - \overline{\bm \phi^{\star}}) (\bm \phi^{\star}_m - \overline{\bm \phi^{\star}})' + k_0 M/(k_0 + M) (\bm m_{\bm \mu} - \overline{\bm \phi^{\star}}) (\bm m_{\bm \mu} - \overline{\bm \phi^{\star}})'$, and $\overline{\bm \phi^{\star}} = \frac{1}{M}\sum_{m = 1}^M \bm \phi^{\star}_m$. When sampling from \eqref{eq:FC_OmegaG_marg}, we resort to the direct sampler by \citep{Lenkoski_2013}, which avoids the computation of the normalizing constant, which is intractable for non-decomposable graphs.

Next, we perform a joint update of $(\bm \Omega_G, G, G_0)$ from their full-conditional. Various algorithms for sampling from the joint distribution of a $G$-Wishart random variable and the underlying graph $G$ have been proposed. Among others, we mention the reversible jump approach for decomposable graphs \citep{Giudici_Green1999}, and the accept/reject method by \cite{Carvalho_etal2007} applicable to the more general case of non-decomposable graphs. Here, we follow \cite{mohammadi2015bayesian} who develop an algorithm based on a birth-and-death process to propose the addition and the removal of the edges of the graph $G_0$ -- and consequently in our model, the addition and the removal of a clique in $G$. The advantage of this method is that every move is accepted, allowing for a more efficient exploration of the sample space. We present in the following how to extend the existing methodology to deal with both $G$ and $G_0$.

\paragraph{Update via birth-and-death process}
For notational convenience, in what follows we omit the conditioning on hyper-parameters lower down in the hierarchy of the model which do not appear in the full conditional under consideration. 

Consider the graphs $G_0 = (V_0, E_0)$ and $G = (V, E)$ introduced in Section 2.3 of the main manuscript. Let $\mathcal{G}_0$ be the sample spaces over which $G_0$ is defined, i.e. the space of all the possible graphs defined over the set of edges $V_0$. Similarly, let $\mathcal{G}$ be the sample space for the graph $G$, obtained by applying the bijection $f: \mathcal{G}_0 \rightarrow \mathcal{G}$, such that $\mathcal{G} = f(\mathcal{G}_0)$. We point out that $\mathcal{G}$ is a subset of the usual sample space for graph, i.e. the space of all the possible graphs defined over the set of edges $V$. Let $(\bm \Omega_g, G, G_0)$ be the current state of the MCMC chain. We consider the following continuous-time birth-and-death Markov process:
\begin{itemize}
	\item[\textbf{Death:}] Each edge $e_0 = (h,k) \in E_0$ dies independently from the others according to a Poisson process with rate $\delta_{e_0}(\bm \Omega_G, G, G_0 \mid \bm \phi^{\star})$. The overall death rate is $\delta(\bm \Omega_G, G, G_0 \mid \bm \phi^{\star}) = \sum_{e_0 \in E_0} \delta_{e_0}(\bm \Omega_G, G, G_0 \mid \bm \phi^{\star})$. If the death of an edge occurs, the process jumps to the state $(\bm \Omega_{G^{-\bm e_{hk}}}, G^{-\bm e_{hk}}, G_0^{-e_0})$, in which  $G_0^{-e_0} = (V_0, E_0\setminus e_0)$, $G^{-\bm e_{hk}} = f(G_0^{-e_0}) = (V, E\setminus \bm e_{hk})$, and $\bm \Omega_{G^{-\bm e_{hk}}} \in \mathbb{P}_{G^{-\bm e_{hk}}}$. Here, $\mathbb{P}_{G}$ indicates the space of precision matrices associated to the graph $G$, while $ E\setminus \bm e_{hk}$ is the set of edges in $G$ from which we have removed the edges in the clique $\bm e_{hk}$ defined by the edge $e_0 \in E_0$.	

	\item[\textbf{Birth:}] Each edge $e_0 = (h,k) \notin E_0$ is born independently from the others according to a Poisson process with rate $\beta_{e_0}(\bm \Omega_G, G, G_0 \mid \bm \phi^{\star})$. The overall birth rate is $\beta(\bm \Omega_G, G, G_0 \mid \bm \phi^{\star}) = \sum_{e_0 \notin E_0} \beta_{e_0}(\bm \Omega_G, G, G_0 \mid \bm \phi^{\star})$. If the birth of an edge occurs, the process jumps to the state $(\bm \Omega_{G^{+\bm e_{hk}}}, G^{+\bm e_{hk}}, G_0^{+e_0})$, in which $G_0^{+e_0} = (V_0, E_0\cup e_0)$, $G^{+\bm e_{hk}} = f(G_0^{+e_0}) = (V, E\cup \bm e_{hk})$, and $\bm \Omega_{G^{+\bm e_{hk}}} \in \mathbb{P}_{G^{+\bm e_{hk}}}$. Similarly to the death event, $\mathbb{P}_{G}$ indicates the space of precision matrices associated to the graph $G$, while $ E\cup \bm e_{hk}$ is the set of edges in $G$ from which we have added the edges in the clique $\bm e_{hk}$ defined by the edge $e_0 \in E_0\cup e_0$.	
\end{itemize}
The birth and death processes are independent Poisson processes, so the waiting time between two successive events is exponentially distributed with mean $1/(\delta(\bm \Omega_G, G, G_0 \mid \bm \phi^{\star}) + \beta(\bm \Omega_G, G, G_0 \mid \bm \phi^{\star}))$. Therefore, given the current state of the chain, the probability of the next event is:
\begin{itemize}
	\item $P(\mbox{ death of edge } e_0 \mid \bm \phi^{\star}) = \frac{\delta_{e_0}(\bm \Omega_G, G, G_0 \mid \bm \phi^{\star})}{\delta(\bm \Omega_G, G, G_0 \mid \bm \phi^{\star}) + \beta(\bm \Omega_G, G, G_0 \mid \bm \phi^{\star})}, \quad e_0 \in E_0$.
	
	\item $P(\mbox{ birth of edge } e_0 \mid \bm \phi^{\star}) = \frac{\beta_{e_0}(\bm \Omega_G, G, G_0 \mid \bm \phi^{\star})}{\delta(\bm \Omega_G, G, G_0 \mid \bm \phi^{\star}) + \beta(\bm \Omega_G, G, G_0 \mid \bm \phi^{\star})}, \quad e_0 \notin E_0$,
\end{itemize}
A straightforward extension of Theorem \ref{th:Theorem1} in \cite{mohammadi2015bayesian}, reported below, leads to an expression for the birth and death rates in the case of the two graphs $(G_0,G)$, which assures that the stationary distribution of such birth-and-death process is the desired one, i.e. the joint posterior $p(\bm \Omega_G, G, G_0 \mid \bm \phi^{\star})$. After selecting the type of move to perform using the probabilities above, we then generate the new precision matrix from the appropriate full conditional in \eqref{eq:FC_OmegaG_marg}.
\begin{theorem}\label{th:Theorem1}
	The birth-and-death process described above has stationary distribution $p(\bm \Omega_G, G, G_0 \mid \bm \phi^{\star})$ if, for each edge $e_0 = (h,k)$ with $h,k \in V_0$:
	
	\begin{align*}
	&\delta_{e_0}(\bm \Omega_G, G, G_0 \mid \bm \phi^{\star}) p(\bm \Omega_G \setminus (\bm \omega_{hk}, \bm \omega_{kk}), G, G_0 \mid \bm \phi^{\star}) \\
	&= \beta_{e_0}(\bm \Omega_{G^{-\bm e_{hk}}}, G^{-\bm e_{hk}}, G^{-e_0}_0 \mid \bm \phi^{\star}) p(\bm \Omega_{G^{-\bm e_{hk}}} \setminus \bm \omega_{kk}, G^{-\bm e_{hk}}, G^{-e_0}_0 \mid \bm \phi^{\star}),
	\end{align*}
	
	with $\bm e_{hk} = f(e_0 = (h,k))$ and $\bm \omega_{hk} = \{\left[\bm \Omega_G\right]_{ij} \mid (i,j) \in \bm e_{hk} \}$.
\end{theorem}
The proof of the above theorem follows exactly the same steps as the proof in \cite{mohammadi2015bayesian}, by noting that all the conditions that hold for a single edge also hold for a clique.

\subsection{Missing values at first time of observation}
As discussed in Section 2 of the manuscript, missing values of the processes at the first time of observation (i.e., 3 months antenatal) can be imputed by treating them as unknown parameters in the model. Therefore, at each iteration of the MCMC algorithm, we only need to sample a new value of the state of the $h$-th process  at time one from:
\begin{equation*}
	p\left(Y^{(h)}_{i1} = k \mid Y^{(h)}_{i2}, \bm \lambda^{(h)} \right) \propto p^{(h)}_{i2} \left( k, Y^{(h)}_{i2}; \bm \lambda^{(h)}_{i2},\epsilon^{(h)}_{i2}\right) \pi^{(h)}_{i1}\left(k; \bm \lambda^{(h)}_{i1}\right), \quad k \in \{1, \dots, d^{(h)}\}
\end{equation*}
for $i = 1, \dots, N$ and $h = 1, \dots, p$. The computation of the above probabilities is straightforward when the likelihood terms are known in closed form (e.g., when $d^{(h)} = 2$), and can be approximated numerically for problems with a higher number of states.

\subsection{Updating allocations and other hyper-parameters}
The update of the variables $(\bm c, \bm S, \bm \phi)$ follows the steps detailed in \cite{argiento2019infinity}, adapted to our modelling setting.

We report here the steps necessary for updating the other parameters of the model, namely $\bm \beta$, $\bm \gamma$, $\bm \mu$ and the vectors $\bm \phi^\star = \left(\bm \phi^\star_1, \dots, \bm \phi^\star_M\right)$ associated with the $M$ components of the proposed model. The updates described below are performed independently for each process $h = 1, \dots, p$. The updates for non-conjugate parameters in the model are performed following the adaptive Metropolis-Hastings (MH) algorithm for multivariate variable of \cite{haario2001adaptive}, where new candidates are proposed from a Gaussian distribution centred on the value of the parameter at the current iteration, and with covariance matrix equal to an appropriately re-scaled version of the sample covariance matrix obtained using the samples produced so far in the MCMC chain.
\begin{itemize}
	
	\item For each $m = 1, \dots, M$, we have that a-priori $\bm \phi^\star_m \mid \bm \mu, \bm \Omega_G \sim P_0 = N_{D_p}(\bm \mu, \bm \Omega_G)$. We  propose a new vector $\left(\bm \phi^\star_m\right)^{\text{new}}$ from a Gaussian random walk proposal centred on the current value $\bm \phi^\star_m$ and with diagonal covariance matrix with non-zero entries equal to $0.25$. Recall that each vector of subject-specific $\log$-baseline transition rates $\bm \phi_i$, for $i = 1, \dots, N$, is linked to the unique value $\bm \phi^\star_m$ via the allocation variable $c_i = m$. Therefore, proposing a new values of $\bm \phi^\star_m$ has an effect on the likelihood terms involving all the subjects in the $m$-th cluster via the new transition rates $\left( \bm \lambda^{(h)} \right)^{\text{new}} = \left\{ \left(\bm \lambda^{(h)}_i \right)^{\text{new}}: c_i = m \right\}$ for $h = 1, \dots, p$. The move is accepted with probability:
	$$
	\min \left\{ 1, \frac{\prod_{h = 1}^p p\left(\bm Y^{(h)} \mid \left( \bm \lambda^{(h)} \right)^{\text{new}}, \bm \beta^{(h)}, \bm \gamma^{(h)}\right) P_0\left(\left(\bm \phi^\star_m\right)^{\text{new}}\right)}{\prod_{h = 1}^p p\left(\bm Y^{(h)} \mid \bm \lambda^{(h)}, \bm \beta^{(h)}, \bm \gamma^{(h)}\right) P_0\left(\bm \phi^\star_m\right)} \right\}
	$$
		
	\item The prior for $\bm \beta^{(h)}$  is $ \text{MN}_{g^{(h)} \times d^{(h)}(d^{(h)} - 1)}(\bm 0, \bm U_{\bm \beta^{(h)}}, \bm V_{\bm \beta^{(h)}})$, which is not conjugate. Therefore, we use an adaptive MH step for the update of its vectorized form $vec(\bm \beta^{(h)})$, for $h = 1, \dots, p$. The acceptance probability is:
	$$
	\min \left\{ 1, \frac{\prod_{h = 1}^p p\left(\bm Y^{(h)} \mid \bm \lambda^{(h)}, \left(\bm \beta^{(h)}\right)^{\text{new}}, \bm \gamma^{(h)}\right) p\left(\left(\bm \beta^{(h)}\right)^{\text{new}}\right)}{\prod_{h = 1}^p p\left(\bm Y^{(h)} \mid \bm \lambda^{(h)}, \bm \beta^{(h)}, \bm \gamma^{(h)}\right) p\left(\bm \beta^{(h)}\right)} \right\}
	$$
	
	\item The prior for $\bm \gamma^{(h)} $ is $ \text{MN}_{q^{(h)} \times d^{(h)}(d^{(h)} - 1)}(\bm 0, \bm U_{\bm \gamma^{(h)}}, \bm V_{\bm \gamma^{(h)}})$, which is not conjugate. Therefore, we use an adaptive MH step for the update of its vectorized form $vec(\bm \gamma^{(h)})$, for $h = 1, \dots, p$. The acceptance probability is:
	$$
	\min \left\{ 1, \frac{\prod_{h = 1}^p p\left(\bm Y^{(h)} \mid \bm \lambda^{(h)}, \bm \beta^{(h)}, \left(\bm \gamma^{(h)}\right)^{\text{new}}\right) p\left(\left(\bm \gamma^{(h)}\right)^{\text{new}}\right)}{\prod_{h = 1}^p p\left(\bm Y^{(h)} \mid \bm \lambda^{(h)}, \bm \beta^{(h)}, \bm \gamma^{(h)}\right) p\left(\bm \gamma^{(h)}\right)} \right\}
	$$
	
	\item We assumed $\bm \mu \mid \bm \Omega_G \sim \text{N}_{D_p}(\bm m_{\bm \mu}, \bm \Omega_G)$, and therefore:
	$$\bm \mu \mid \bm \phi, \bm \Omega_G \sim \text{N}_{D_p} \left( \frac{\sum_{m = 1}^M \bm \phi_m + k_0 \bm m_{\bm \mu}}{M + k_0}, (M + k_0) \bm \Omega_G \right)
	$$
	
	\item We assume a priori $k_0 \sim \text{Gamma}(a_{k_0}, b_{k_0})$, and therefore:
	$$
	k_0 \mid \bm \mu, \bm \Omega_G \sim \text{Gamma} \left( a_{k_0} + D_p/2, b_{k_0} + \frac{1}{2} (\bm \mu - \bm m_{\bm \mu})' \bm \Omega_G (\bm \mu - \bm m_{\bm \mu}) \right)
	$$
	
\end{itemize}

\clearpage
\section{Sensitivity analysis on $\Lambda$ and $\gamma_S$}\label{SMsec:Sensitvity}
We present a sensitivity analysis used in support of the choice of the hyperparameters $\Lambda$ and $\gamma_S$, appearing in the definition of the mixture part of the proposed model (see model (3) in the main manuscript). In particular, $\Lambda$ directly influences the distribution of the number of components in the mixture model, while $\gamma_S$ relates to the prior distribution of the corresponding unnormalized weights. For each combination of values of these hyperparameters such that $\Lambda, \gamma_S \in \{0.01, 0.1, 1, 10\}$, we run the MCMC algorithm for 50000 iterations after an initial brun-in period of 1000 iterations used to initialize the adaptive components of the MCMC. The last 10000 iterations are saved and thinned every second iteration, yielding posterior samples of size 5000. Each posterior chain so-obtained is used to compute the posterior mode of the number of clusters $K_N$ and of the number of components $M$, the number of clusters in the estimated Binder's partition, as well as the posterior 95 \% credibility interval for the entropy of the sampled partitions. We report in Table \ref{tab:Sensitivity} such values for each combination of the hyperparameter $\Lambda$ and $\gamma_S$. In general, an increase in the value of $\Lambda$ produces partitions characterized by a higher number of components, increasing both the number of allocated (i.e., of clusters) and non-allocated ones. On the other hand, increasing the value of $\gamma_S$ has a coarsening effect on the resulting partition, reducing the number of components to three even when $\Lambda = 10$. Notice how the value $\gamma_s = 10$ induces partitions with less variability, as depicted by the same values explored by the Entropy function. In the application of Section 3 in the main manuscript, we present detailed results corresponding to the combination $(\Lambda, \gamma_S) = (0.01, 0.1)$.

\begin{table}[!ht]
	\caption[]{GUSTO cohort study: results from the sensitivity analysis on the hyperparameters $\Lambda$ and $\gamma_S$ driving the distribution of the partition induced by the mixture model. We summarize the explored partitions with the posterior mode of the number of clusters $K_N$ and components $M$, the number of clusters in the estimated Binder's partition (top part of each cell), and 95 \% credible intervals for the entropy (bottom part of each cell).}
	\label{tab:Sensitivity}
	\begin{tabular}{c|cccc}
		\diagbox[width=10em]{$\Lambda$}{$\gamma_S$} & 0.01 & 0.1 & 1 & 10 \\\hline
		0.01 & \makecell{(3,3,3) \\ (0.95, 1.00)} & \makecell{(3,3,3) \\ (0.96, 1.01)} & \makecell{(4,4,4) \\ (1.30, 1.41)} & \makecell{(3,3,3) \\ (0.97, 1.01)}\\\hline
		0.1 & \makecell{(4,4,4) \\ (1.30, 1.35)} & \makecell{(5,5,5) \\ (1.51, 1.65)} & \makecell{(4,4,4) \\ (1.29, 1.37)} & \makecell{(3,3,3) \\ (0.97, 1.01)}\\\hline
		1 & \makecell{(5,6,5) \\ (1.49, 1.63)} & \makecell{(5,5,7) \\ (1.50, 1.71)} & \makecell{(7,7,6) \\ (1.62, 1.97)} & \makecell{(3,3,3) \\ (0.97, 1.01)}\\\hline
		10 & \makecell{(6,14,5) \\ (1.52, 1.80)} & \makecell{(9,15,10) \\ (1.64, 2.11)} & \makecell{(13,13,19) \\ (1.94, 2.68)} & \makecell{(3,3,3) \\ (0.97, 1.01)}\\
	\end{tabular}
\end{table}

\pagebreak
\section{Comparison with alternative models}
We provide in this section a simulated example to assess the performance of the proposed model in comparison to existing alternatives. In particular, we fit four different competing models to the simulated dataset: the proposed SUMS model where the distribution of the parameters $\bm \phi$ is (i) a mixture with random number of components as in model (6) (main text); (ii) a Dirichlet Process mixture with centring measure $P_0 = \text{N}_{D_p}(\bm \mu, \bm \Omega_G)$ (DP-SUMS); (iii) a parametric distribution equal to $P_0$ (Param-SUMS) or (iv) the model for disease progression proposed by \cite{DeIorio_etal_2018}. The latter is devised for data with only one response process (i.e., disease/remission), and allows the inclusion of time-homogeneous, as well as time-varying, covariates. The time-varying covariates are included in the model via the specification of a suitable model for both continuous and categorical ones. The distribution of the covariates, as well as the instantaneous transition rates of the response process, are linked by the specification of a latent time-varying subject-specific health function. Clustering of the subjects is driven by the health function trajectories as well as the subject-specific instantaneous transition rates. 
For fairness, we compare the four model on the basis of common characteristics: the estimate of the partition of the subjects, the effect of the time-homogeneous covariates on the evolution of the response process, and the relevance of the time-varying covariates (both binary and continuous) on disease progression.

The data are simulated for $N = 200$ subjects as follows: we select three binary processes, representing the disease progression process and two binary time-varying covariates. We assume two underlying clusters and sample the corresponding allocation variables from a discrete distribution with proportions $(1/3, 2/3)$ for the two clusters, respectively. The baseline instantaneous transition rates in each cluster are vectors of dimension $D_p = 6$ since each process has only two states. We fix these values using empirical estimates of the transition rates of the observed binary processes in the GUSTO data (i.e., STAI-s, STAI-t and PSQI), obtained via the R package \texttt{msm}. This yields the six instantaneous $\log$-transition rates $\bm \phi^{\text{crude}} = \log(0.12, 0.37, 0.11, 0.26, 0.21, 0.34)$. We set $\bm \phi^{\star, 1} = \bm \phi^{\text{crude}}$ and $\bm \phi^{\star, 2} = \bm \phi^{\text{crude}} + 1$ for the two clusters, respectively. The three multi-state processes are generated independently of each other. 

The times of observation are selected following \cite{DeIorio_etal_2018} where, starting from time zero, the interval $(0,10)$ is split by adding values generated from a normal distribution left truncated at 0.5 with unitary mean and variance. The left truncation is chosen as the smallest time interval observed in the GUSTO dataset (corresponding to half a year). Notice that the simulated time points are the same for all the three processes, as required by data modelled using model (iv) by \cite{DeIorio_etal_2018}. However, the proposed SUMS model allows also for process-specific times of observation.

We simulate one  time-homogeneous covariate  from a standard normal distribution and one from a Bernoulli distribution with success probability 0.25. Two time-varying continuous covariates are simulated from Gaussian distributions with variance equal to 0.5 and means which are functions of the simulated times of observation. The first covariate has a linear trend $t/10$, while the second one has a trigonometric mean $cos(t/10 2 \pi)$. The time-homogeneous and time-varying continuous covariates are included only in the Cox proportional hazard model for the binary disease progression process, and the corresponding regression coefficients are fixed equal to:
$$
\bm \beta =
\begin{bmatrix}
	1 & 1\\
	-1 & -1 &
\end{bmatrix}, \quad
\bm \gamma =
\begin{bmatrix}
	0.75 & -1.25\\
	1.25 & -0.75
\end{bmatrix}
$$

Finally, the data for the multi-state processes are simulated using the R package \texttt{msm} \citep{jackson_2011}. This simulated example is included in the sample GitHub code attached to this work.

The hyperparameter settings when fitting the SUMS model are the same as described in Section 3 (main text). However, the random effect distribution of $\bm \phi$ differs among models. In particular, for the proposed SUMS model we fix $\gamma_S = \Lambda = 0.1$, while for the DP model we select a mass parameters $\alpha = 0.178$ which corresponds to assuming the a-priori expected number of clusters approximately equal to 2. The hyperprior elicitation in the case of model (iv) follows \cite{DeIorio_etal_2018}. We run the MCMC algorithm for each model for a total of 25000 iterations and use the last 5000, after a thinning of 2, for posterior inference.

\begin{figure}[ht]
	\includegraphics[width=0.75\textwidth, angle = -90]{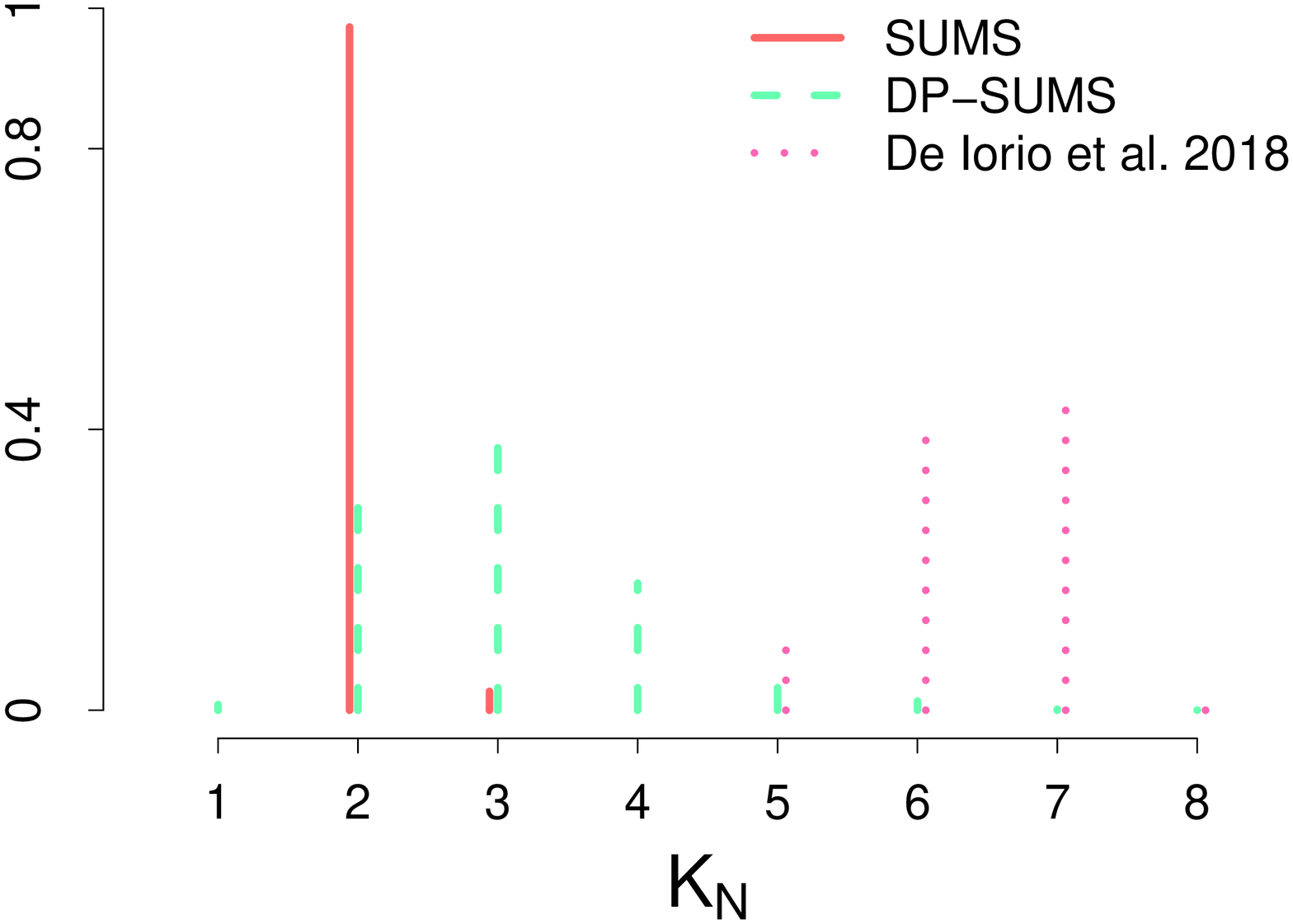}
	\caption{Posterior distribution of the number of clusters for competing models.}
	\label{SMfig:SimulStudy_K_N_hist}
\end{figure}

\begin{figure}[ht]
	\centering
	\subfloat[SUMS]{\includegraphics[width=0.275\textwidth, angle = -90]{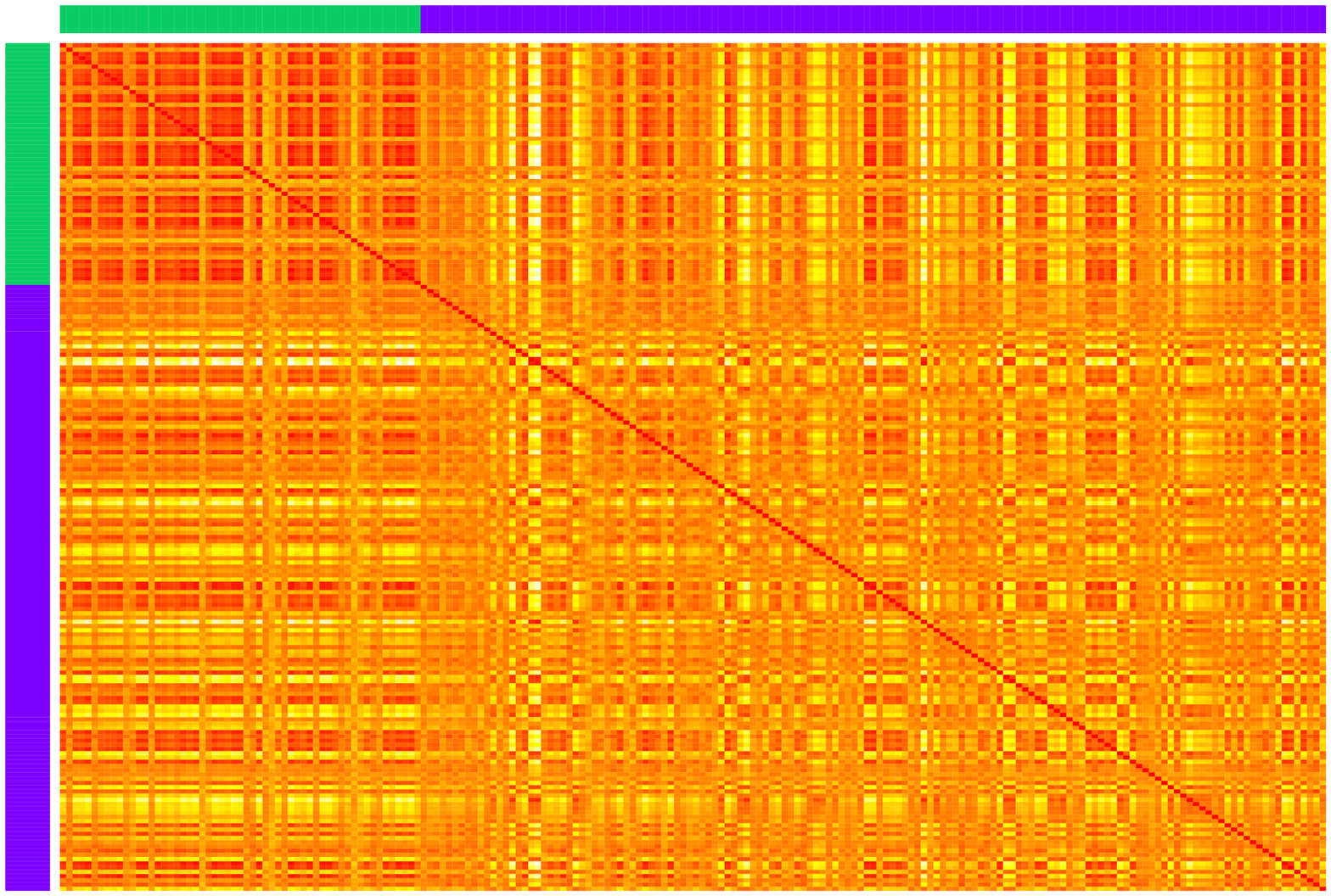}}
	\subfloat[DP-SUMS]{\includegraphics[width=0.275\textwidth, angle = -90]{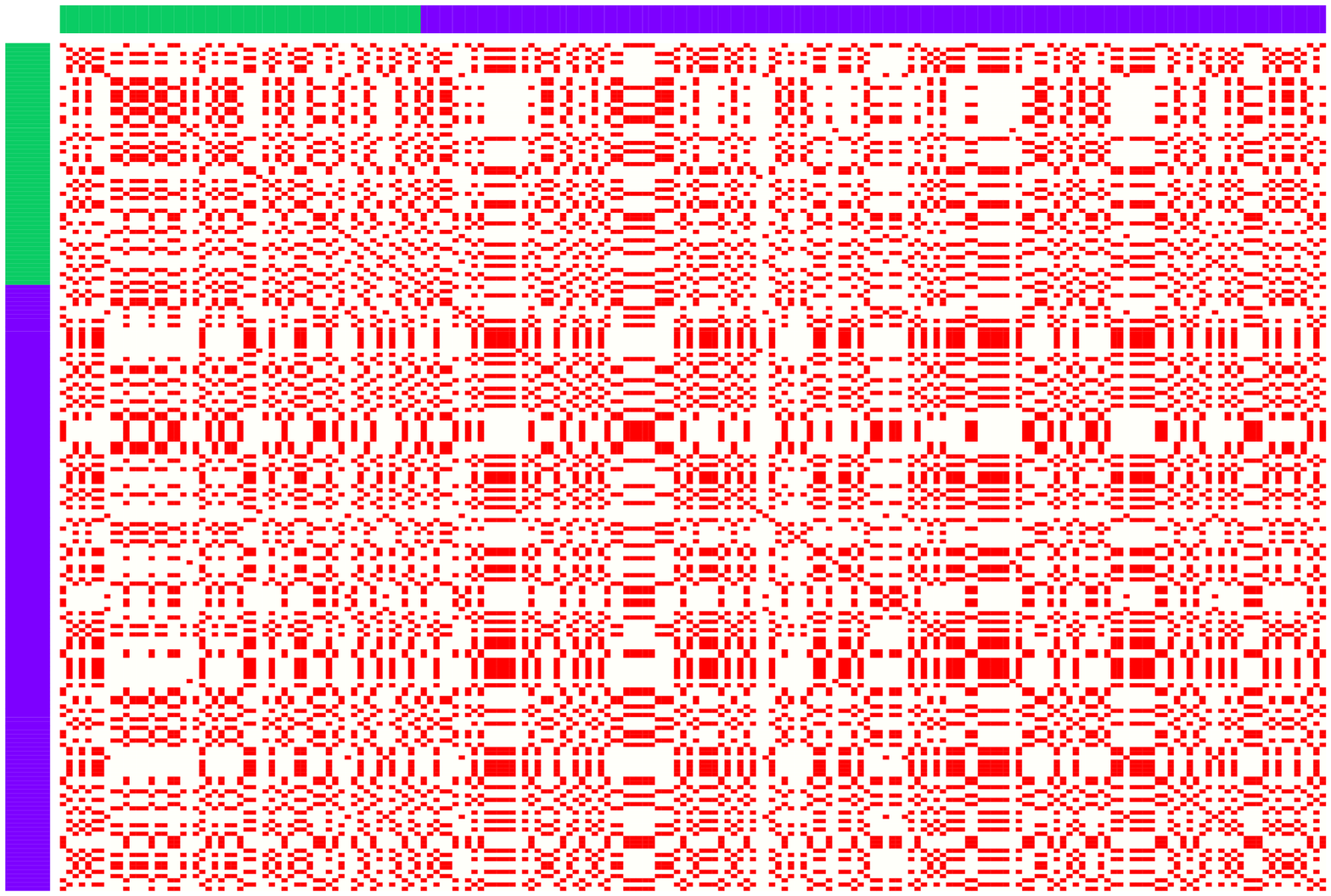}}
	\subfloat[\cite{DeIorio_etal_2018}]{\includegraphics[width=0.275\textwidth, angle = -90]{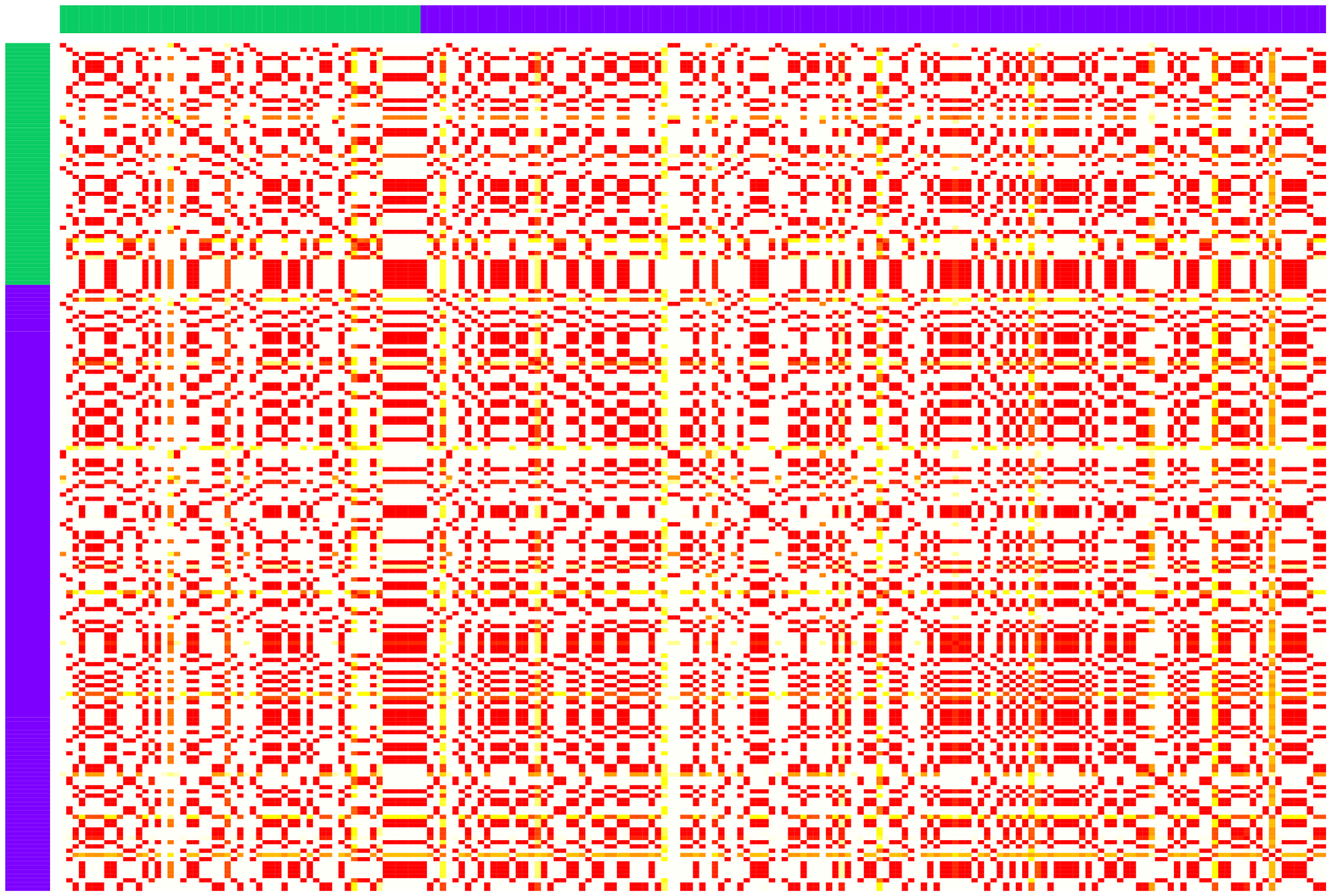}}
	\caption{Posterior co-clustering probability for competing models. The colour bars at the edges represent the true clustering of the subjects, identified by different colours.}
	\label{SMfig:SimulStudy_CoClusteringProbs}
\end{figure}

Firstly, we show inference on the partition of the subjects, which is obtainable with the proposed and competing models, excluding the parametric version of the SUMS model. Figures \ref{SMfig:SimulStudy_K_N_hist} and \ref{SMfig:SimulStudy_CoClusteringProbs} display the posterior distributions of the number of clusters and the posterior co-clustering probability matrices, respectively. We notice worse performance of the DP-SUMS and of the \citep{DeIorio_etal_2018} models. 


\begin{figure}[ht]
	\includegraphics[width=0.75\textwidth, angle = -90]{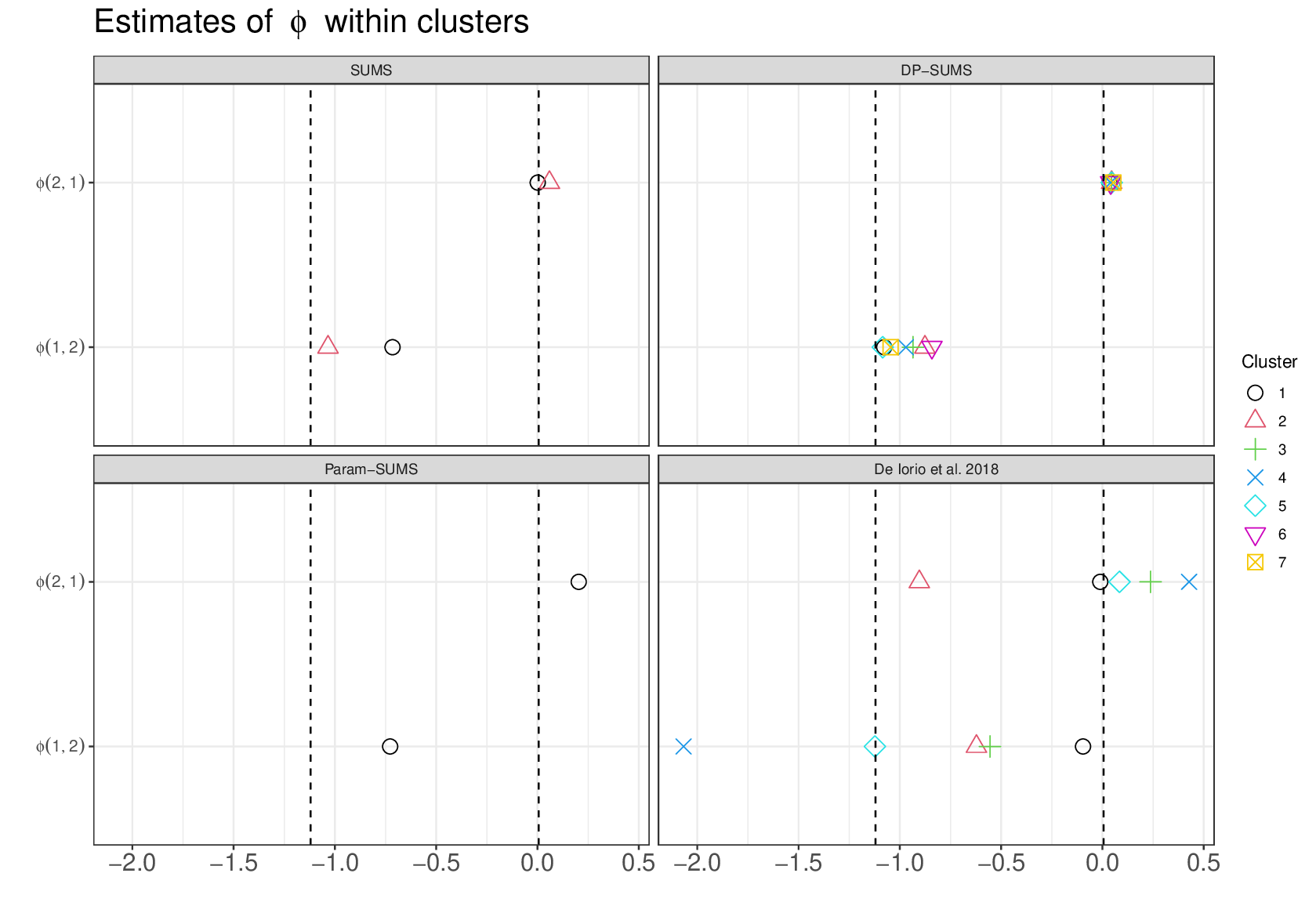}
	\caption{Posterior mean of the empirical average of the instantaneous $\log$-baseline transition rates within each cluster (values averaged over the subjects). Each panel corresponds to a different model. Within each panel, the dots represent different cluster estimates, identified by different shapes/colours. The values of the instantaneous $\log$-baseline transition rates used to simulate the data are indicated by the dashed vertical lines.}
	\label{SMfig:SimulStudy_phi_inClusters_all}
\end{figure}

Figure \ref{SMfig:SimulStudy_phi_inClusters_all} shows the posterior mean of the average of the subject-specific $\log$-baseline transition rates within each cluster for each model. In the figure, we only show the results relative to the binary response process, which is common to all models under comparison. In these estimates, the partitions are fixed to the ones obtained minimizing the Binder loss function \citep{Binder_1978}. Different symbols and colours correspond to different clusters. Notice how most of the estimates lie close to the two true unique values of $\bm \phi$ (dashed lines), apart from the parametric model which is recovering only one of the two values. In particular, the DP-SUMS model overestimates the number of clusters (estimated equal to 7), while the model of \cite{DeIorio_etal_2018} shows higher variability in the estimation of the unique values of the $\log$-transition rates.

\begin{figure}[ht]
	\includegraphics[width=0.75\textwidth, angle = -90]{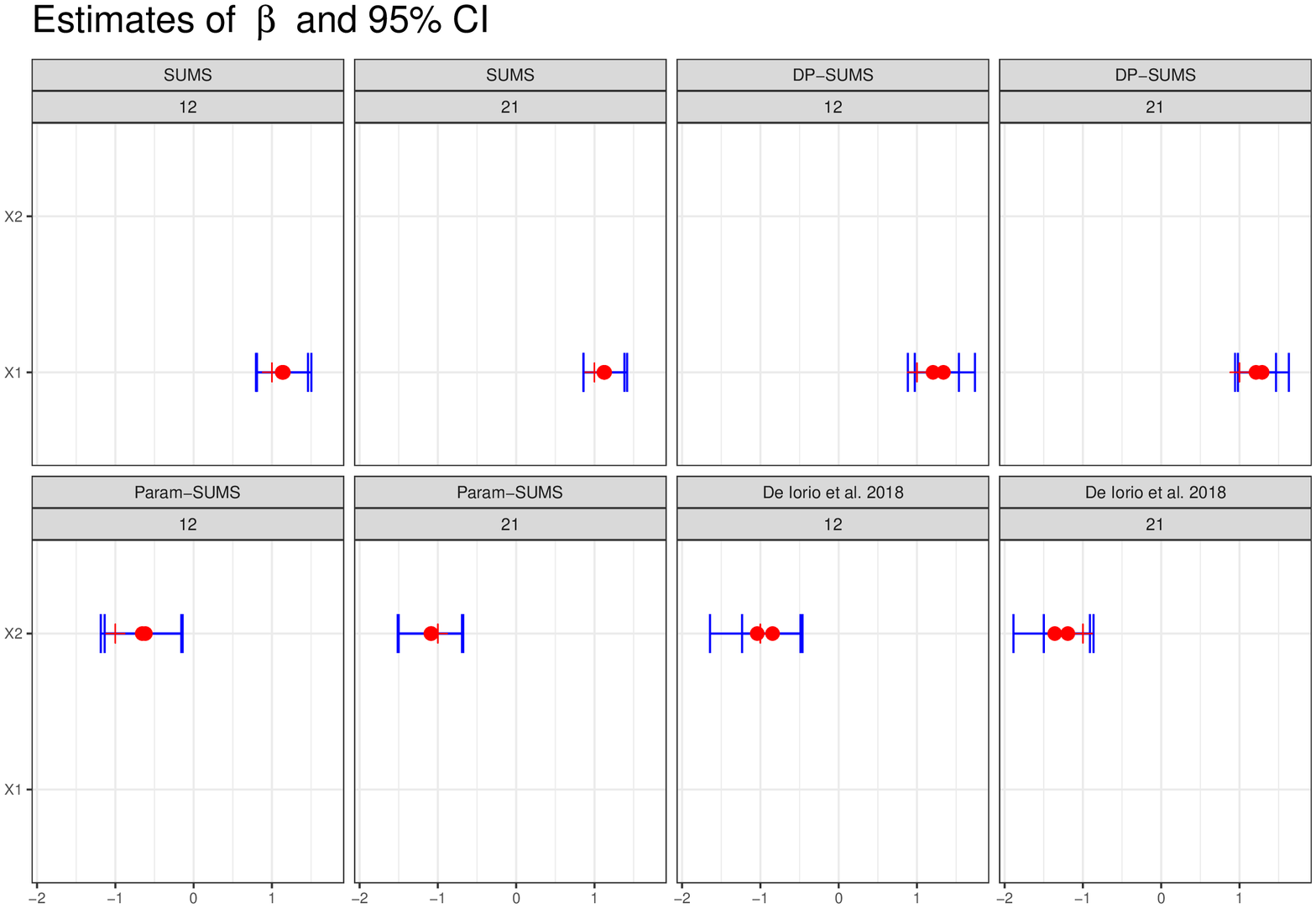}
	\caption{Posterior mean and 95\% CI of the regression coefficients $\bm \beta$ associated with the time-homogeneous covariates. Each panel corresponds to a different model and possible transition (from state 1 to state 2 and vice-versa). The true values used to simulate the data are depicted as red crosses.}
	\label{SMfig:SimulStudy_beta_CI_all}
\end{figure}

We compute the posterior mean and 95\% credible intervals (CI) of the regression coefficients $\bm \beta$, associated to the time-homogeneous covariates $\bm X$ in the Cox regression model for the transition rates. These parameters are included in all the four models and therefore can be fully compared. All the models are able to recover the true value of the coefficients, since the true values are always included in the 95\% CI of the corresponding parameters (see Figure \ref{SMfig:SimulStudy_beta_CI_all}). The same is true for the time-varying continuous covariates $\bm Z$. Indeed, the proposed model, as well as its DP and parametric alternatives, recover the corresponding regression coefficients $\bm \gamma$ correctly. The disease progression model of \cite{DeIorio_etal_2018} does not include such covariates in the same way into the model, but instead models these terms as Gaussian distributions whose mean is a linear function of the latent health parameter $\theta_{ij}$, namely $\mathbb{E}[Z_l] = c_{0l} + c_{1l} \theta_{ij}$, for each covariate $l = 1, 2$. The slope coefficient $c_{1l}$ in this model indicates whether there is a dependency of the health status on the $l$-th covariate, which can be therefore interpreted as a relevant effect of such covariate on the evolution of the subjects' disease status over time. We find that, in accordance with the simulation setting, the 95\% CI's of these coefficients do not contain the value zero, and therefore correctly estimate their effects.

\clearpage
\section{Additional Tables}\label{secSM:Additional_Tables}

\begin{table}[!ht]
	\caption[]{GUSTO cohort study: thresholds used for the discretization of the mental health scores. In brackets are indicated the labels used for each state. The last column reports the observation times in years and the percentage of missing values at the first time of observation (in brackets). The pre-natal time points refer to 6 months into pregnancy (3 months before birth).}
	\label{tabSM:scores_cat}
	\hspace{-1cm}
	\begin{tabular}{c|cccccc|c}
		& \multicolumn{6}{c}{\textbf{Thresholds}} & \textbf{Time-points} (years) \\\hline
		\makecell{\textbf{BDI} \\ category} & \multicolumn{2}{c}{\makecell{$0 \leq x < 14$ \\ Low (1)}} & \multicolumn{2}{c}{\makecell{$14 \leq x < 20$ \\ Mild (2)}} & \multicolumn{2}{c|}{\makecell{$x \geq 20$ \\ Depressed (3)}} & \makecell{pre-natal: 0.5 (5.15 \%) \\ \makecell{post-natal: 0.25, 1, 2, \\ 3, 4.5, 6, 8.5}} \\
		\makecell{\textbf{EPDS} \\ category} & \multicolumn{2}{c}{\makecell{$0 \leq x < 5$ \\ Low (1)}} & \multicolumn{2}{c}{\makecell{$5 \leq x < 9$ \\ Mild (2)}} & \multicolumn{2}{c|}{\makecell{$x \geq 9$ \\ Clinical (3)}} & \makecell{pre-natal: 0.5 (1.47 \%) \\ post-natal: 0.25, 2} \\
		\makecell{\textbf{STAI-s} \\ category} & \multicolumn{3}{c}{\makecell{$20 \leq x < 40$ \\ non-Clinical (1)}} & \multicolumn{3}{c|}{\makecell{$x \geq 40$ \\ Clinical (2)}} & \makecell{pre-natal: 0.5 (2.94 \%) \\ \makecell{post-natal: 0.25, 1, 2, \\ 3, 4.5, 6, 8.5}}\\
		\makecell{\textbf{STAI-t} \\ category} & \multicolumn{3}{c}{\makecell{$20 \leq x < 40$ \\ non-Clinical (1)}} & \multicolumn{3}{c|}{\makecell{$x \geq 40$ \\ Clinical (2)}} & \makecell{pre-natal: 0.5 (3.68 \%) \\ \makecell{post-natal: 0.25, 1, 2, \\ 3, 4.5, 8.5}} \\
		\makecell{\textbf{PSQI} \\ quality of sleep} & \multicolumn{3}{c}{\makecell{$0 \leq x < 6$ \\ good (1)}} & \multicolumn{3}{c|}{\makecell{$x \geq 6$ \\ poor (2)}} & \makecell{pre-natal: 0.5 (37.13 \%) \\ \makecell{post-natal: 0.25, 0.5, 1, 2, \\ 3, 4.5}} \\
	\end{tabular}
\end{table}

\begin{table}[!ht]
	\caption[]{GUSTO cohort study: time-homogeneous covariates used in the analysis. The second column indicates the levels or ranges of observation for the categorical or continuous covariates, respectively. The third column indicates the percentage of missing information, imputed with the R package \texttt{mice}.}
	\label{tabSM:covariates}
	\hspace{-1cm}
	\begin{tabular}{c|c|c}
		\textbf{Variable name} & \textbf{Levels/Range} & \textbf{Missing} \\\hline
		
		mother ethnicity & \makecell{1 = Chinese, 2 = Malay, 3 = Indian} & 0.00 \% \\\cline{2-2}

		mother highest education & \makecell{1 = Below Secondary,\\ 2 = Above Secondary and below University,\\ 3 = University and above} & 0.00 \% \\\cline{2-2}

		marital status & {1 = married, 2 = not married} & 0.00 \% \\\cline{2-2}
						
		mother age at delivery & $\mathbb{R}^+$ & 0.00 \% \\\cline{2-2}           

		age at first menstrual cycle & $\mathbb{R}^+$ & 3.32 \% \\\cline{2-2}

		Gender of the baby & {1 = Female, 2 = Male} & 0.00 \% \\\cline{2-2}
						
		smoking exposure pre-pregn. & {1 = no, 2 = yes} & 3.65 \% \\\cline{2-2}

		alcohol consumption pre-pregn. & {1 = no, 2 = yes} & 0.66 \% \\\cline{2-2}
		
		general health & {1 = bad, 2 = fair, 3 = good} & 0.00 \% \\\cline{2-2}    
		
		stress affected health & 1 = none, 2 = moderately, 3 = extremely & 4.31 \% \\\cline{2-2}
		
		Cotinine [ng/mL] & $\mathbb{R}^+$ & 3.65 \% \\\cline{2-2}

		BFI (6 dim.) & $\mathbb{R}^+$ & 0.00 \% \\\cline{2-2}
		
		MCA & $\mathbb{R}^+$ & 1.00 \% \\\hline
		
	\end{tabular}
\end{table}

\clearpage
\section{Additional Figures}\label{secSM:Additional_Figures}

We include in this section some additional figures, discussed in the application Section 3 of the main manuscript.

\begin{figure}[ht]
	\centering
	\includegraphics[angle = -90, width=0.45\textwidth]{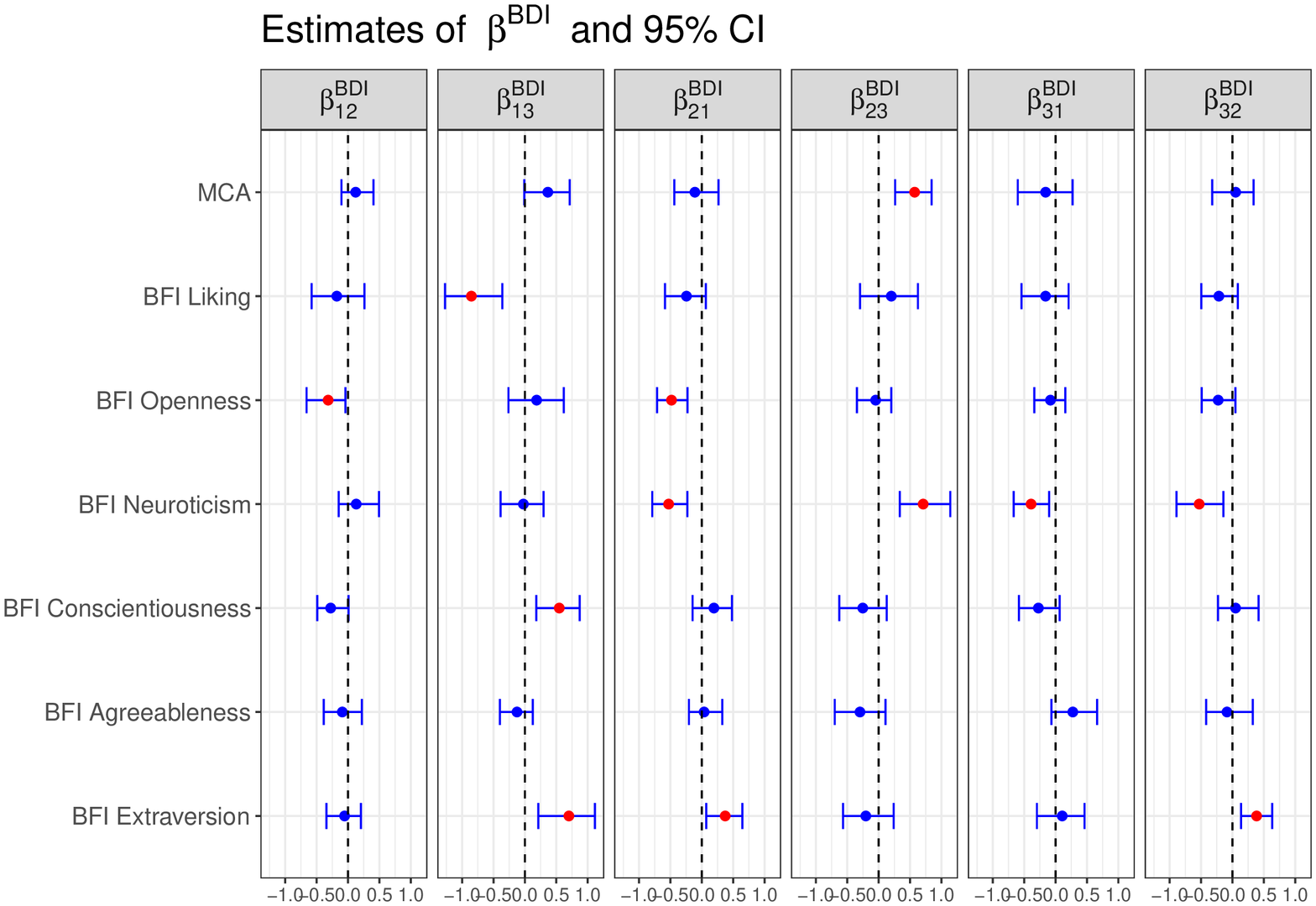}
	\includegraphics[angle = -90, width=0.45\textwidth]{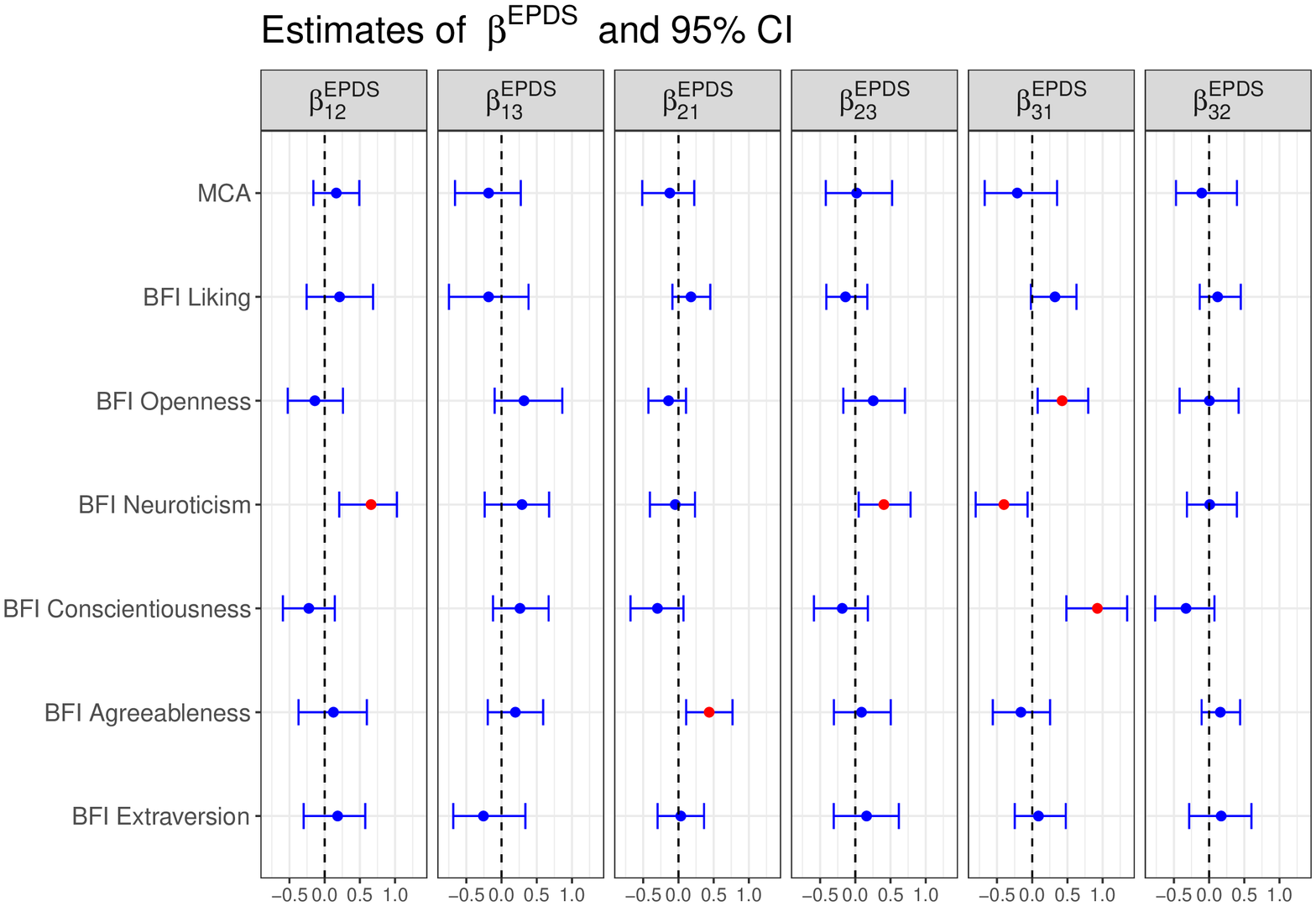}\\
	\includegraphics[angle = -90, width=0.45\textwidth]{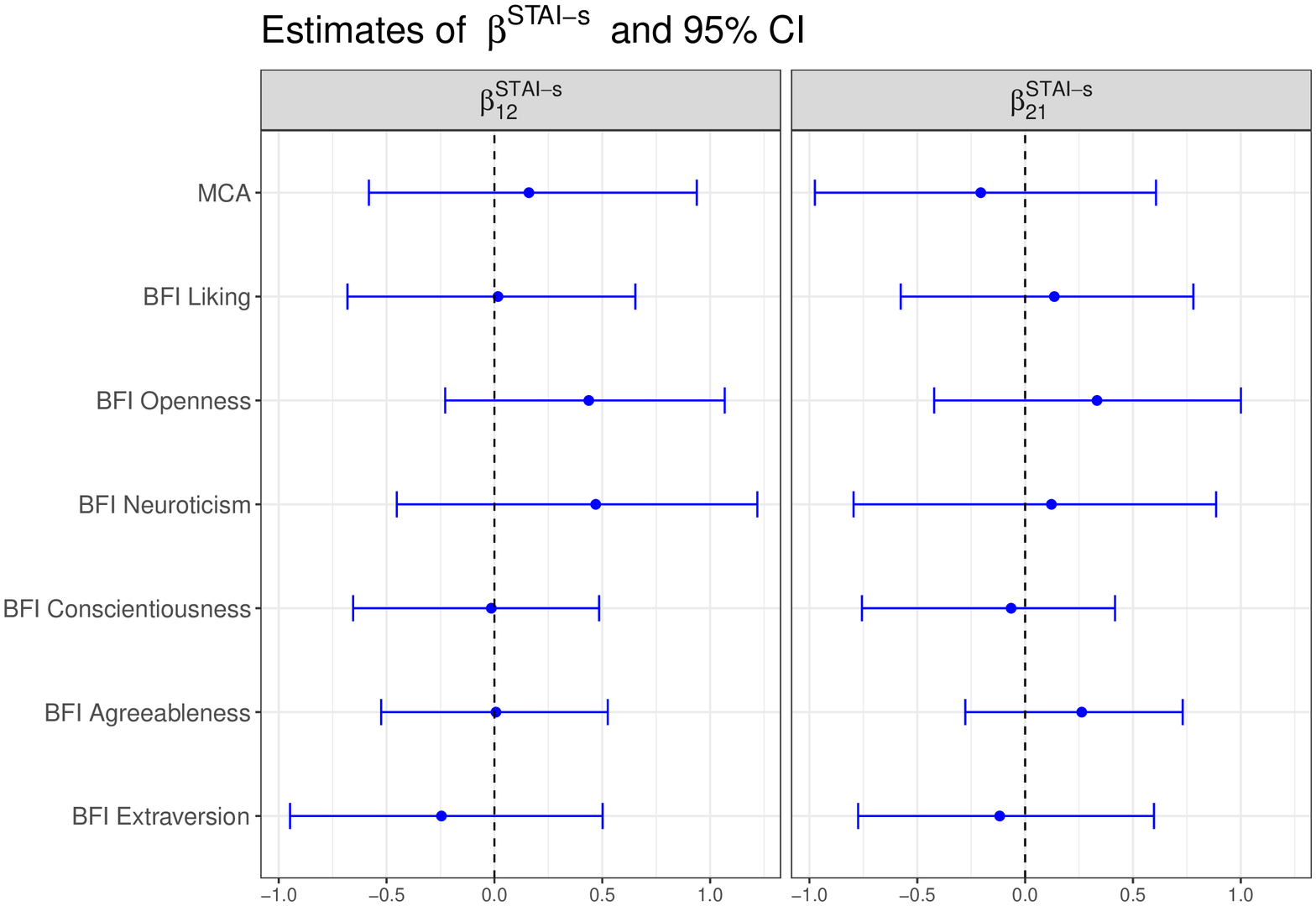}
	\includegraphics[angle = -90, width=0.45\textwidth]{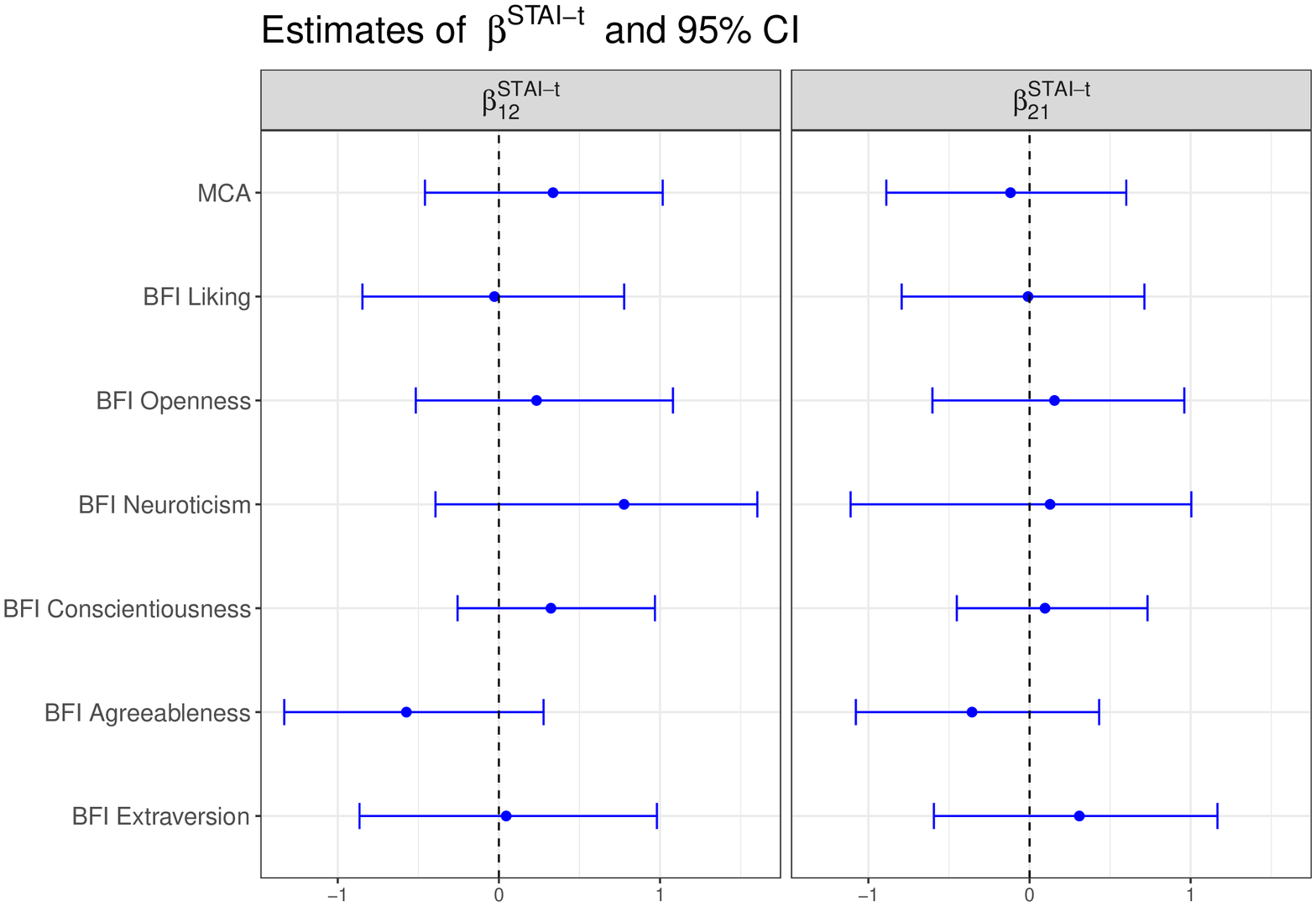}
	\caption{Posterior means and 95\% credible intervals for the coefficients $\bm \beta^{(h)}$, for $h = 1, \dots, 4$ relative to the BFI and MCA scores. Within each panel, each column refers to a different transition for the considered process, while each row is referred to a different BFI dimension or to the MCA score. The red dots indicate a relevant effect of the covariate on the corresponding transition (i.e., the 95\% credible interval does not contain the value 0).}
	\label{SMfig:Beta_BFI}
\end{figure}

Figure \ref{SMfig:Beta_BFI} shows the posterior distribution of the regression coefficients $\bm \beta^{(h)}$ relative to the BFI, CTQ and PBI covariates. In each figure, the posterior estimates (posterior 95\% credible intervals and posterior means) are grouped based on the mental health process they refer to. The posterior mean of the regression coefficients is plotted in red colour whenever the 95\% credible interval does not include the value zero, indicating that the corresponding covariate is relevant for the description of the transition probabilities of the selected process.

\begin{figure}[ht]
	\subfloat[]{\includegraphics[width=0.3\textwidth, angle = -90]{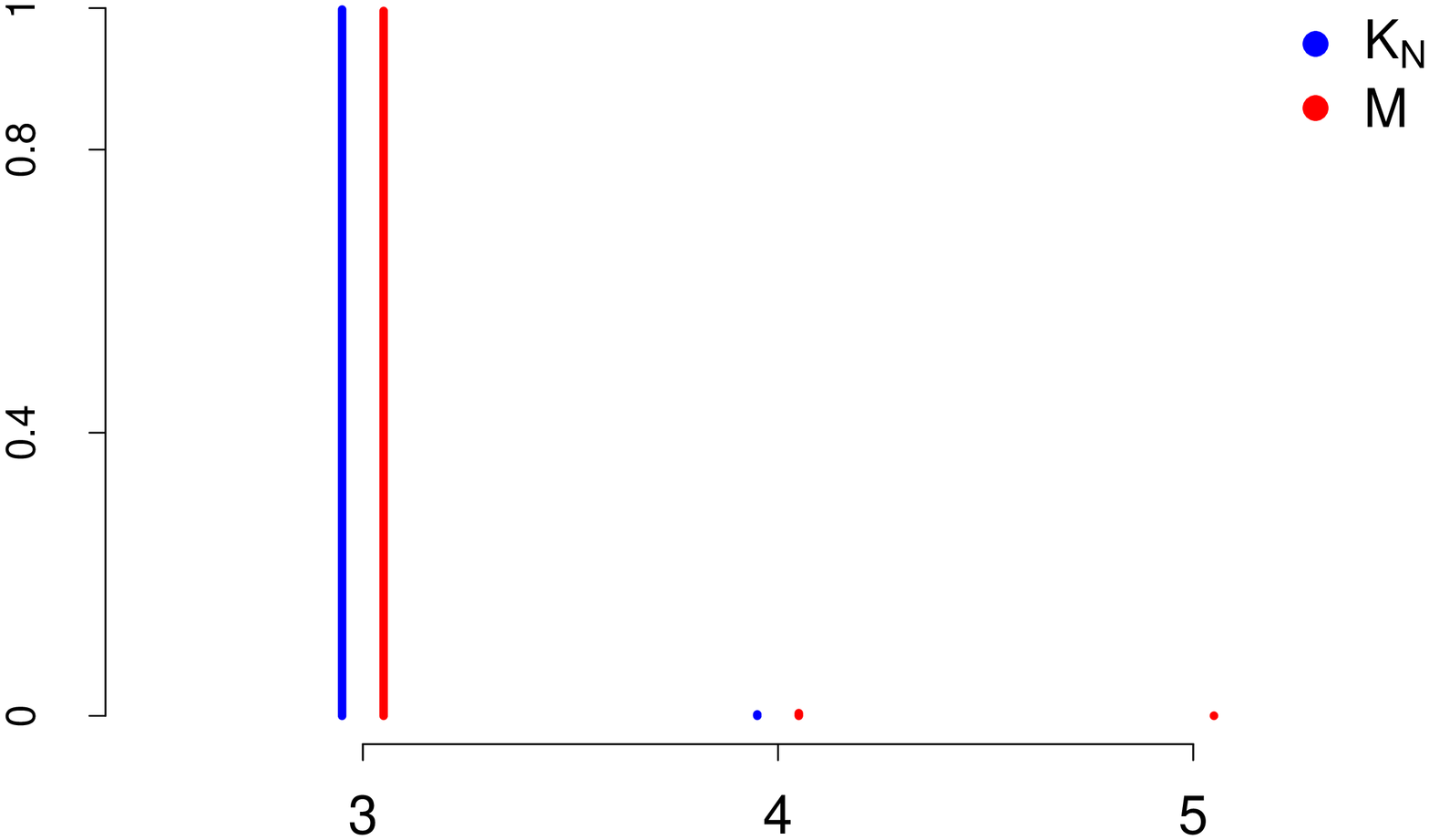}}
	\subfloat[]{\includegraphics[width=0.35\textwidth, angle = -90]{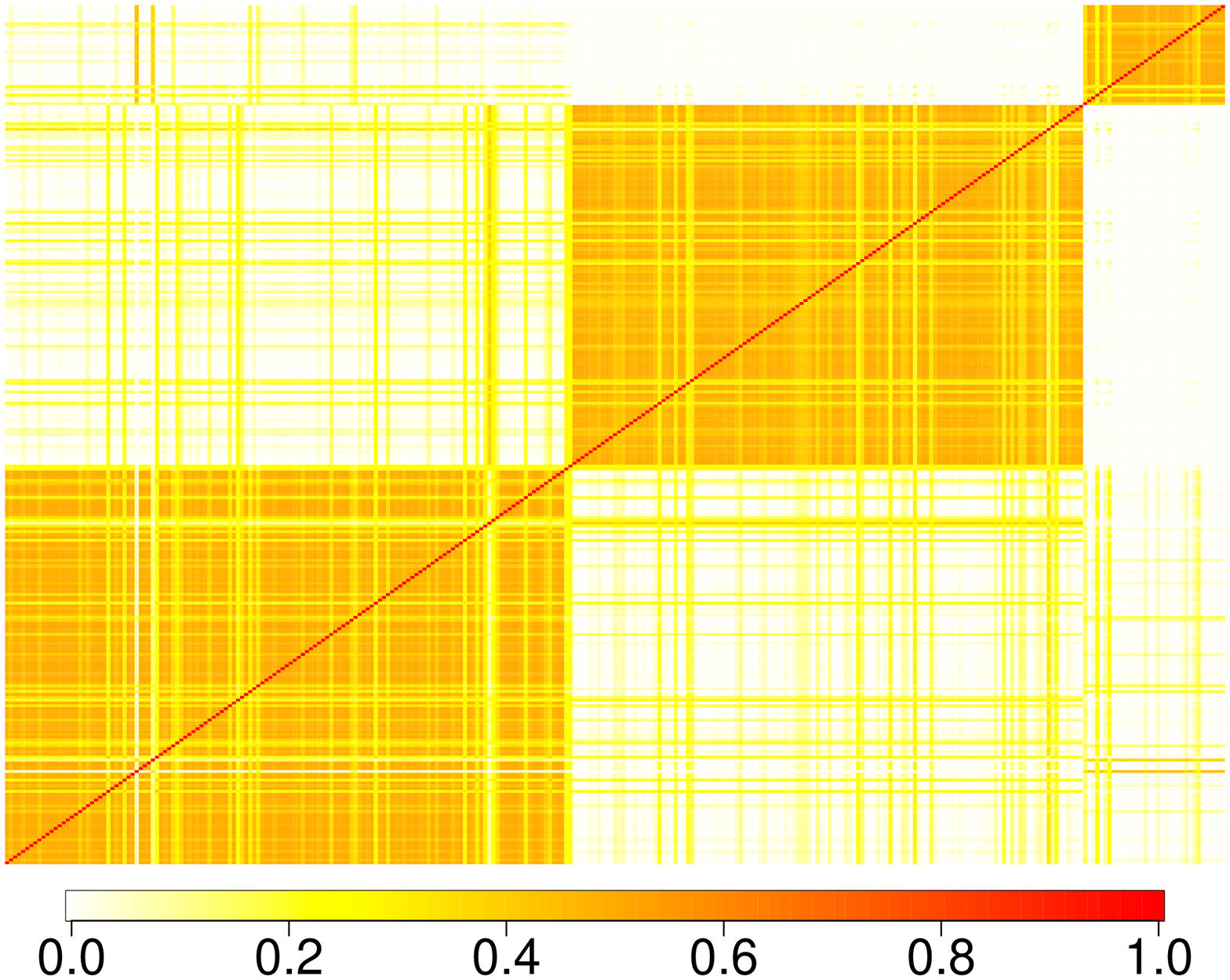}}
	\caption{Posterior distribution of \textbf{(a)}: the number of clusters $K_N$ and of the number of components $M$ in the mixture. \textbf{(b)} presents the posterior co-clustering probabilities for the $N = 301$ subjects.}
	\label{SMfig:KN_M_PPI}
\end{figure}

Figure \ref{SMfig:KN_M_PPI} includes the posterior distribution of the number of clusters, components, and the posterior pairwise clustering probabilities.

\section{Clustering a-posteriori}

We discuss in this section the results obtained on the clustering of the subjects. 

\begin{figure}[ht]
	\includegraphics[width=0.75\textwidth, angle = -90]{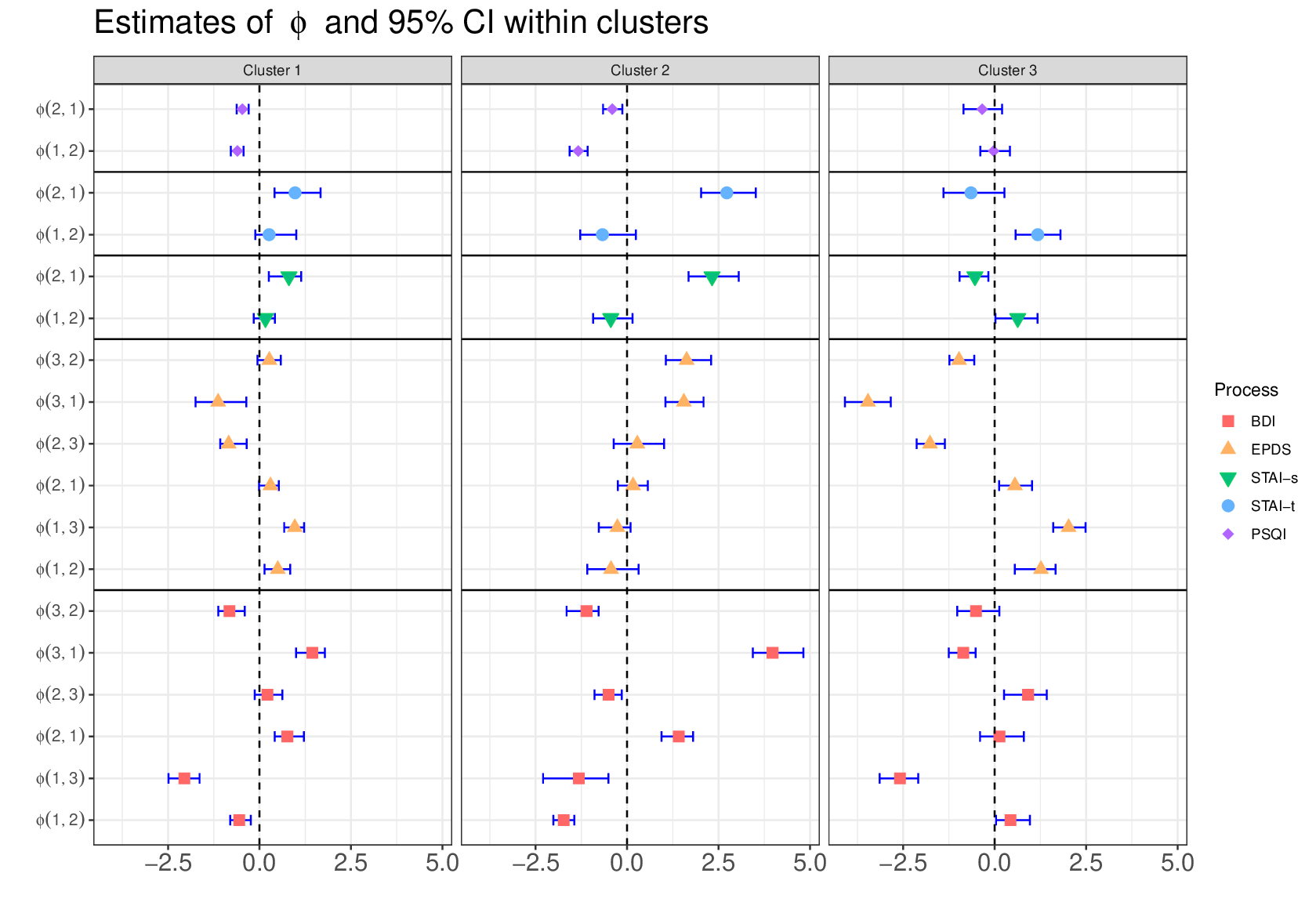}
	\caption{Posterior means and 95\% credible intervals of the instantaneous $\log$-transition rates $\phi^{(h)}(r,s)$ for each process $h = 1, \dots, p_0$. The vertical dashed lines represent the value 0, while the horizontal continuous lines divide the estimates for the five processes. The estimates are obtained by fixing the partition of the subjects to the Binder's partition, and re-running the algorithm for the conditional model. Each sub-plot refers to one of the clusters in the fixed partition.}
	\label{SMfig:phi_inClusters}
\end{figure}

First, we estimated a partition of the subjects a-posteriori using the Binder loss function method \citep{Binder_1978}. Thus, we run the MCMC algorithm conditionally tot he estimated partition, i.e. without updating the partition labels $\bm c$. This yielded a posterior chain for the parameters of the model conditionally to the estimated Binder partition. We report in Figure \ref{SMfig:phi_inClusters} the posterior distribution of $\phi^{(h)}(r,s)$ so-obtained, for $r \rightarrow s$ and $h = 1, \dots, p$. Cluster-specific estimates of transition rates differ among clusters. For instance, the instantaneous transition rates for process BDI are lower for transitions to states corresponding to worse mental health outcomes with respect to the opposite transition in Clusters 1 and 2, when compared to those in Cluster 3. Similarly, transition rates corresponding to improvement in the BDI score are higher in Clusters 1 and 2 rather than Cluster 3. An analogous behavior can be observed for the EPDS process. In particular, in Cluster 3, the transition rates from state 3 (Clinical) to state 1 (Low) or state 2 (Mild) are lower than in the other two clusters, suggesting that the subjects in this cluster are less prone to improvement in their mental health status. The processes with only two states (STAI-s, STAI-t and PSQI) also seem to present differences between clusters. For example, Clusters 1 and 2 are characterized by lower transition rates for the processes STAI-s and STAI-t from state 1 (non-Clinical) to state 2 (Clinical), and vice-versa for Clusters 3. Analogously, the PSQI process shows higher transition rates from state 1 (good sleep quality) to state 2 (poor sleep quality) in Clusters 3. 

\clearpage
\bibliographystyle{plainnat}
\bibliography{Biblio}